\def\version{January 9, 2002}

\documentclass[reqno,11pt]{amsart}

\usepackage{amssymb, epsf}

\def\be{\begin{equation}}
\def\ba{\begin{align}}
\def\bm{\begin{multline}}
\def\bfig{\begin{figure}[htb]}
\def\efig{\end{figure}}

\newcommand{\bibit}[1]{\bibitem[#1]{#1}}
\newcommand{\paper}[1]{{\it #1}, }
\newcommand{\journal}[4]{#1 {\bf #2}, #3 (#4)}
\newcommand{\CMP}{Commun.\ Math.\ Phys.}
\newcommand{\HPA}{Helv.\ Phys.\ Acta}
\newcommand{\JPA}{J.\ Phys.\ A}
\newcommand{\JSP}{J.\ Stat.\ Phys.}

\newcommand{\PRB}{Phys.\ Rev.\ B}
\newcommand{\PRL}{Phys.\ Rev.\ Lett.}
\newcommand{\RMP}{Rev.\ Math.\ Phys.}

\numberwithin{equation}{section}
\newtheorem{theorem}{Theorem}[section]
\newtheorem{proposition}[theorem]{Proposition}
\newtheorem{lemma}[theorem]{Lemma}
\newtheorem{corollary}[theorem]{Corollary}

\newcounter{eqs}
\newcommand{\startalpheqno}{\setcounter{eqs}{\value{equation}}
\setcounter{equation}{0} \addtocounter{eqs}{1}
\renewcommand{\theequation}{\arabic{section}.\arabic{eqs}{\it \alph{equation}}}
}
\newcommand{\stopalpheqno}{\setcounter{equation}{\value{eqs}}
\renewcommand{\theequation}{\arabic{section}.\arabic{equation}}}

\newcommand{\nn}{\nonumber}
\def\bbbone{{\mathchoice {\rm 1\mskip-4mu l} {\rm 1\mskip-4mu l} {\rm
1\mskip-4.5mu l} {\rm 1\mskip-5mu l}}}

\renewcommand{\leq}{\;\leqslant\;}
\renewcommand{\geq}{\;\geqslant\;}

\newcommand{\dd}{{\rm d}}
\newcommand{\e}[1]{\,{\rm e}^{#1}\,}
\newcommand{\ii}{{\rm i}}
\newcommand{\sumtwo}[2]{\sum_{\substack{#1 \\ #2}}}

\newcommand{\inftwo}[2]{\inf_{\substack{#1 \\ #2}}}

\def\Tr{{\operatorname{Tr\,}}}

\def\dist{{\operatorname{dist\,}}}

\newcommand{\expval}[1]{\langle #1 \rangle}
\newcommand{\bigexpval}[1]{\bigl\langle #1 \bigr\rangle}

\newcommand{\compl}{{\text{\rm c}}}

\newcommand{\upchi}{\raise1pt\hbox{$\chi$}}
\newcommand{\charfct}[1]{\,\upchi\, \bigl[ \text{$ #1 $} \bigr]}
\newcommand{\const}{{\text{\rm const}}}

\def\writefig#1 #2 #3 {\rlap{\kern #1 truecm \raise #2 truecm
\hbox{#3}}}
\def\figtext#1{\smash{\hbox{#1}} \vspace{-5mm}}

\newcommand{\caB}{{\mathcal B}}

\newcommand{\caQ}{{\mathcal Q}}
\newcommand{\caR}{{\mathcal R}}

\newcommand{\bbR}{{\mathbb R}}

\newcommand{\bbZ}{{\mathbb Z}}

\begin{document}

{\small
\hfill
\version
}
\vspace{2mm}

\title{Segregation in the Falicov-Kimball model}

\author{James K. Freericks, Elliott H. Lieb, Daniel Ueltschi}

\address{James K. Freericks,
Department of Physics,
Georgetown University,
Washington, DC 20057, USA,
{\rm freericks@physics.georgetown.edu,
http://www.physics.georgetown.edu/$\sim$jkf}}

\address{Elliott H. Lieb,
Departments of Mathematics and Physics,
Princeton University,
Jadwin Hall,
Princeton, NJ 08544, USA;
{\rm lieb@math.princeton.edu,
http://www.math.princeton.edu/$\sim$lieb}}

\address{Daniel Ueltschi,
Department of Physics,
Princeton University,
Jadwin Hall,
Princeton, NJ 08544, USA.
{\it Present address:} Department of Mathematics,
University of California,
Davis, CA 95616, USA;
{\rm ueltschi@math.ucdavis.edu,
http://math.ucdavis.edu/$\sim$ueltschi}}

\maketitle

\renewcommand{\thefootnote}{}
\footnote{J. K. F. is partially supported by the Office of Naval Research under grant
N00014-99-1-0328. E. H. L. and D. U. are partially supported by the National Science
Foundation under grant PHY-98 20650.}
\footnote{\copyright{} 2001 by the authors. This paper may be reproduced,
in its entirety, for non-commercial purposes.}

\setcounter{footnote}{0}
\renewcommand{\thefootnote}{\arabic{footnote}}

\begin{quote}
{\small
{\bf Abstract.}
The Falicov-Kimball model is a simple quantum lattice model that
describes light and heavy electrons interacting with an on-site repulsion; alternatively,
it is a model of itinerant electrons and fixed nuclei.
It can be seen as a simplification of the Hubbard model; by neglecting the kinetic
(hopping) energy of the spin up particles, one gets the Falicov-Kimball model.

We show that away from half-filling, i.e.\ if the sum of the densities of both kinds
of particles differs from 1, the particles segregate at zero temperature and for
large enough repulsion. In the language of the Hubbard model, this
means creating two regions with a positive and a negative magnetization.

Our key mathematical results are lower and upper bounds for the sum of the lowest eigenvalues of the discrete
Laplace operator in an arbitrary domain, with Dirichlet boundary
conditions. The lower bound consists of a bulk term, independent
of the shape of the domain, and of a term proportional to the
boundary. Therefore, one lowers the kinetic energy of the itinerant
particles by choosing a domain with a small boundary. For the Falicov-
Kimball model, this corresponds to having a single `compact' domain
that has no heavy particles.
}

\vspace{1mm}
\noindent
{\footnotesize {\it Keywords:} Falicov-Kimball model, segregation, phase separation.

\noindent
{\it 2000 Math.\ Subj.\ Class.:} 82B10, 82B20, 82B26.}
\end{quote}

\vspace{5mm}

{\footnotesize \tableofcontents}

\section{Introduction}

\subsection{The Falicov-Kimball model}

Introduced thirty years ago to describe the semicon\-ductor-metal transition in SmB$_6$ and
related materials \cite{FK}, the Falicov-Kimball model is a simple lattice model with rich
and interesting properties. The system consists of two species of spinless electrons with different
effective masses: one species has infinite mass (so the particles do not move --- we call
them `classical particles'), while the second species represents itinerant
spinless electrons whose kinetic energy is represented by a hopping matrix. The Hamiltonian in a
finite domain $\Omega \subset \bbZ^d$ is
\be
\label{defHamFK}
H_\Omega^U(\{w_x\}) = -\sumtwo{x,y \in \Omega}{|x-y|=1} c_x^\dagger c_y + 2d \sum_{x \in
\Omega} n_x + U \sum_{x \in \Omega} w_x n_x.
\end{equation}
Here, $c_x^\dagger$, $c_x$, denote creation, annihilation operators of an electron at site $x$; $n_x
= c_x^\dagger c_x$; $w_x=0,1$ is the number of classical particles (`heavy electrons') at
$x$, and $U \geq 0$ is an on-site repulsion between the two species of particles.
$H_\Omega^U(\{w_x\})$ represents the energy of the electrons under a potential $U w_x$.
The term $2d \sum n_x$ in \eqref{defHamFK} is for convenience only. It makes $H^U_\Omega$
positive, and this term only adds $2d$ times the electron number, $N$. At zero
temperature, one is typically interested in the configurations of classical particles that
minimize the ground state energy of the electrons.

The model was reinvented in \cite{KL} as a simplification of the Hubbard model, by
neglecting the hoppings of electrons of spin $\uparrow$, say. This simplification changes
the nature of the model somewhat, mainly because the continuous SU(2) symmetry
is lost. Connections between the two models are therefore not
immediate; however, the greater knowledge obtained for the Falicov-Kimball model may help
in understanding the Hubbard model.

Rigorous results in \cite{KL} include a proof that equilibrium states display
long-range order of the chessboard type when both species of particles have
density 1/2; this holds for all dimensions greater than 1 and for {\it all} $U\neq0$
(including $U<0$), provided the temperature is low enough. The model is reflection
positive under a suitable magnetic field, or when the electrons are replaced by
hard-core bosons; this property can be used to establish long-range order \cite{MP}.
Perturbative methods allow for an extension of these results for large $U$ and small
temperature, see \cite{LM, MM, DFF}. Absence of
long-range order when the inverse temperature $\beta$ is small, or $\beta U$ is small, was
also established in \cite{KL}.

One may increase the density of one species and decrease the density of the other species while
maintaining the half-filling condition, namely that the total density is 1. (However, as
was shown in \cite{KL}, the lowest energy is achieved when both species have density 1/2.)
The one-dimensional case was considered in \cite{Lem};
if classical particles and electrons have respective densities $\frac pq$ and $1-\frac pq$,
the ground state is the `most homogeneous configuration' for $U$ large enough; this configuration is
periodic with a period no greater than $q$. Away from half-filling the particles segregate:
classical particles occupy one side of the chain, leaving room for electrons on the other
side. There are several results in 2D. Gruber et.\ al.\
\cite{GJL} performed a $1/U$ expansion and found periodic phases when the density of the
classical particles is $1/3$, $1/4$, $1/5$. This was made rigorous by Kennedy
\cite{Ken}. These results are reviewed in \cite{GM}. The knowledge of the 2D phase diagram
for large $U$ was further extended in
\cite{Ken2, Hal, HK}. New ground states for various rational densities were uncovered; for some
densities they are periodic, but there are also densities where coexistence of
configurations with different periods have minimum energy. The results are summarized in
Fig.\ 1 in \cite{HK}. Finally, it was understood in \cite{DMN} that the 111 interface
is stable, due to the effective interactions between the particles.

\subsection{Away from half-filling}

The purpose of our study is to explore the situation away from half-filling; i.e., we take the
total density to differ from 1. 
For any density away from half-filling we prove that the ground state is
segregated for $U$ large enough. When $U = \infty$, the ground state is
segregated for all densities (at half filling all configurations have
same energy, including segregated and periodic ones).
Hole-particle symmetries for both species of particles
\cite{KL} imply that the results for positive $U$ and densities $(n_{\rm e}, n_{\rm
c})$ of electrons and heavy (classical) particles, transpose to (a) positive $U$ and
densities $(1-n_{\rm e}, 1-n_{\rm c})$, and (b) negative $U$ and densities
$(n_{\rm e}, 1-n_{\rm c})$ or $(1-n_{\rm e}, n_{\rm c})$. For simplicity, we
take the total density $n_{\rm e} + n_{\rm c}$ to be strictly less than 1.

We start our study by taking the limit $U \to \infty$. Electrons are described by wave
functions that vanish on sites occupied by the classical particles, and the question is to find
the arrangement of classical particles that minimizes the energy of the electrons. This amounts to
minimizing the sum of the lowest eigenvalues of the discrete Laplace operator with
Dirichlet boundary conditions. This is explained in Section \ref{seclb}, where it is shown
that the energy per site of $N$ electrons in a finite domain $\Lambda \subset \bbZ^d$ with
volume $|\Lambda|$, is bounded
below by the energy per site of the electrons in the infinite lattice with density $n=N/|\Lambda|$.

One can refine this lower bound by including a term proportional to $|\partial\Lambda|$, the volume of the
boundary $\partial\Lambda$ of $\Lambda$
(Section \ref{seclbb}). This implies that the configuration of the heavy, fixed electrons
that minimizes the ground state energy of the movable electrons has, more or less, one
large hole with relatively small perimeter. Thus the movable particles are separated from
the fixed ones. This behavior was conjectured in \cite{FF} and is opposite to the checkerboard configuration, in which both
kinds of particles are inextricably mixed. Segregation was shown to occur in the ground
state of the $d=1$ model in \cite{Lem}, and of the $d=\infty$ model in \cite{FGM}. The present
paper proves that this holds for all dimensions, and in particular for the relevant
physical situations $d=2$ and $d=3$.

Segregation is more difficult to understand on a heuristic level than the chessboard
phase. The latter is a {\it local} phenomenon that results from effective interactions
between nearest neighbor sites, while the former is a {\it global} phenomenon involving
extended wave functions. This remark
should also apply to the Hubbard model, for which antiferromagnetism is much better
understood than ferromagnetism.

The fact that the sum of the lowest $N$ eigenvalues of  the Laplacian
in a domain of volume $|\Lambda|$ is bounded below by the infinite
volume value at the same density is not unexpected and holds also in
the continuum. Indeed, the original idea, due to Li and Yau \cite{LY}
(see also \cite{LL} Sections 12.3 and 12.11), was demonstrated in the continuum,
and we only adapted it to the lattice context. However, the fact that
the error term is proportional to $|\partial\Lambda|$, the area of the
boundary, is a completely different story. Its proof, at least the one
given here, is complicated. More to the point, such a bound {\it does
not hold in the continuum}. One can easily construct a continuum domain
with finite  volume $|\Lambda|$, but with $|\partial\Lambda|= \infty$,
and for which all eigenvalues are finite.

Taking $U$ large instead of infinite decreases the energy, but the
gain is at most proportional to $|\partial\Lambda|$, as explained in Section \ref{secfinU}.
Therefore the infinite--$U$ segregation effect outweighs this gain and the particles are
still separated (at least when the heavy particle density is far enough
from 1/2).
Finally, an upper bound derived in Section \ref{secub} shows that the energy of electrons
in $\Lambda$ really consists of a bulk term independent of the shape of $\Lambda$, plus a term of the
order of the boundary.

These results are summarized in Theorem \ref{mainthm} below. One can use them to discuss
the electronic free energy at inverse temperature $\beta$, for a {\it fixed} configuration
of classical particles, see Theorem \ref{thmpostemp}. The conclusion of this paper involves a
discussion of first-order phase transitions at finite temperature, of what happens when
classical particles have a small hopping term, and of the possible links with ferromagnetism in
systems of interacting electrons with spins.

In order to present the main result of this paper, we need a few definitions. For $k \in
(-\pi,\pi]^d$, we set
\be
\label{defeps}
\varepsilon_k = 2d - 2\sum_{i=1}^d \cos k_i.
\end{equation}
The energy per site $e(n)$ of a density $n$ of free electrons in the infinite volume
$\bbZ^d$ is
\be
\label{defbulken}
e(n) = \frac1{(2\pi)^d} \int_{\varepsilon_k < \varepsilon_{\rm F}}
\varepsilon_k \, \dd k,
\end{equation}
where the Fermi level $\varepsilon_{\rm F} = \varepsilon_{\rm F}(n)$ is defined by the
equation
\be
n = \frac1{(2\pi)^d} \int_{\varepsilon_k < \varepsilon_{\rm F}} \dd k.
\end{equation}

We can specify the configuration $(w_x)_{x \in \bbZ^d}$ of classical particles by the
domain $\Lambda
\subset \bbZ^d$ consisting of those sites {\it without} particles (holes), that is, $w(x)=1$ if $x \notin \Lambda$ and $w(x)=0$
if $x \in \Lambda$. Let $h_\Lambda^U$ denote the one-particle Hamiltonian whose action on a
square summable, complex function
$\varphi$ on $\bbZ^d$ is
\be
[h_\Lambda^U \varphi](x) = -\sum_{y, |y-x|=1} \varphi(y) + 2d \varphi(x) + U
\upchi_{\Lambda^\compl}(x) \varphi(x).
\end{equation}
Here, $\upchi_{\Lambda^\compl}(x)$ is the characteristic function that is 1 if $x$ belongs
to the complement $\Lambda^\compl$ of $\Lambda$, and is 0 if
$x\notin\Lambda^\compl$.
We define $E_{\Lambda,N}^U$ to be the ground state energy of $N$ electrons for the
configurations defined by $\Lambda$, i.e.\
\be
E_{\Lambda,N}^U = \inf_{\{\varphi_1, \dots, \varphi_N\}} \sum_{i=1}^N (\varphi_i, h_\Lambda^U \,
\varphi_i),
\end{equation}
where the infimum is taken over $N$ orthonormal functions, i.e.\ $(\varphi_i,\varphi_j) =
\delta_{ij}$.
There exist normalized minimizers if the Fermi level is below $U$; they
are not identically zero inside $\Lambda$, and decay exponentially outside.

Notice that $E_{\Lambda,N}^U$ is increasing in $U$, since $(\varphi, h_\Lambda^U \varphi)$
is increasing in $U$ for any $\varphi$.

We define the boundary by $\partial \Lambda = \{ x \in \Lambda : \dist(x,\Lambda^\compl)=1 \}$, where
$\Lambda^\compl$ is the complement of $\Lambda$, i.e., the points in $\bbZ^d$ not in
$\Lambda$. The following theorem
summarizes the results obtained in this paper. It contains upper {\it and} lower bounds for
the ground state energy. We set $n = N/|\Lambda|$.

\begin{theorem}
\label{mainthm}
There are functions $\alpha(n) > 0$ and $\gamma(U)$ with $\lim_{U\to\infty} U
\gamma(U) = 8d^2$, such that for all finite domains $\Lambda$,
$$
\bigl( 2dn-e(n) \bigr) |\partial \Lambda| \geq E_{\Lambda,N}^U - |\Lambda| e(n) \geq \bigl(
\alpha(n) - \gamma(U) \bigr) |\partial \Lambda|.
$$
Furthermore, $\alpha(n) = \alpha(1-n)$, and for $n\leq |S_d|/(4\pi)^d$, it can be
chosen as
$$
\alpha(n) = \frac{2^{d-3}}{\pi^d d^3 |S_d|^{2/d}} n^{1+\frac2d}.
$$
Here, $|S_d|$ is the volume of the unit sphere in $d$ dimensions.
\end{theorem}

An explicit expression for $\gamma(U)$ can be found in Proposition \ref{propfinU}.
The lower bound $\alpha(n)$ vanishes when $n=0$ (no itinerant electrons) or $n=1$
(fully occupied lower band). Theorem \ref{mainthm} is relevant only when $\alpha(n) > \gamma(U)$,
that is, sufficiently away from half-filling (depending on $U$).
The theorem states that the `good' configurations $\Lambda$ for which electrons have low energy
must have small boundaries. As a consequence, the system displays phase separation in the
ground state.

The upper bound is symmetric under the transformation $n \mapsto 1-n$ due to a
symmetry of the Hamiltonian, and it is saturated for
$U=\infty$ by configurations with isolated holes. Indeed, in this case the eigenstates
consist of $\delta$ functions on the holes, with eigenvalues equal to $2d$, and $\partial\Lambda =
\Lambda$.

The lower bound is first explained in Section \ref{seclb} for $U=\infty$ and without the term involving the
boundary. The latter requires more effort and is derived using Lemmas
\ref{lemdiagby}--\ref{lemmu} in Section \ref{seclbb}. Proposition \ref{propfinU} then extends it to
the case of finite $U$. The upper bound is proved in Section \ref{secub}.

Theorem \ref{mainthm} is described in \cite{FLU}, which also reviews the rigorous
results obtained so far for the Falicov-Kimball model.

\subsection{Electrons at low temperature}

It is natural to consider the situation at positive temperature. The relevant object is the
Gibbs state obtained by averaging over the configurations of classical particles, and by
taking the trace of the Gibbs operator $\exp\bigl\{-\beta H_\Omega^U(\{w_x\}) \bigr\}$. We expect the system to
display a first-order phase transition in the grand-canonical ensemble; densities of both
types of particles should have discontinuities as functions of the chemical potentials. But
a rigorous treatment of this phase transition is beyond reach at present. However, we do obtain some
properties of the system when the configuration of the classical particles is fixed, and the electrons
are at positive temperature. Namely, one can extend the estimates of the ground state
energy to estimates of the electronic free energy. The results are described in this
section, and their derivation can be found in Section \ref{secpostemp}.

Let us consider a box $\Omega$ with periodic boundary conditions. The configuration of
classical particles is specified by the set of holes $\Lambda \subset \Omega$ (later, in Corollary
\ref{corIsing}, we shall average over $\Lambda$). With $\mu$ being the chemical potential,
the grand-canonical {\it electronic} free energy (equal to $-|\Omega|/\beta$ times
the pressure) is
\be
\label{deffen}
F_{\Omega,\Lambda}^U(\beta,\mu) = -\frac1\beta \log \Tr \exp \bigl\{ -\beta
H_{\Omega,\Lambda}^U + \beta\mu N_\Omega \bigr\}.
\end{equation}
Here, $H_{\Omega,\Lambda}^U = H_\Omega^U(\{w_x\})$ as defined in \eqref{defHamFK}, $N_\Omega
= \sum_{x\in\Omega} n_x$ is the number of electrons in $\Omega$, and the trace is in the
Fock space of antisymmetric wave functions on $\Omega$.

A simple `guess' for $F_{\Omega,\Lambda}^U$ is obtained by considering independent
electrons, which are either in $\Lambda$ or else in $\Lambda^\compl$. In the latter case the
effective chemical potential is $\mu-U$. Our `guess' would then be
\be
\label{guess}
F_{\Omega,\Lambda}^U(\beta,\mu) \approx |\Lambda| f(\beta,\mu) + (|\Omega|-|\Lambda|)
f(\beta,\mu-U),
\end{equation}
where $f(\beta,\mu)$ is the free energy per site for free electrons:
\be
\label{thermofen}
f(\beta,\mu) = -\frac1\beta \frac1{(2\pi)^d} \int_{[-\pi,\pi]^d} \dd k \, \log\bigl( 1
+ \e{-\beta(\varepsilon_k-\mu)} \bigr).
\end{equation}

Formula \eqref{guess} is, indeed, correct when $U$ is large --- in the sense that the error
is proportional only to $|\partial\Lambda|$. More precisely,

\begin{theorem}
\label{thmpostemp}
There are functions $\bar\alpha(\beta,\mu)>0$ with $\lim_{\beta\to\infty}
\bar\alpha(\beta,\mu) > 0$ if $0<\mu<4d$, and $\bar\gamma(U)$ with $\lim_{U\to\infty}
U \bar\gamma(U) = 16d^2 + 2^{d+3} d^4$, such that for all finite domains $\Omega$ and
$\Lambda\subset\Omega$,
\ba
C_{d,\mu} |\partial\Lambda| + C_{d,\mu}' |\Omega|^{1-\frac1d} &\geq F_{\Omega,\Lambda}^U(\beta,\mu) - \bigl\{ |\Lambda| f(\beta,\mu) + (|\Omega|-|\Lambda|) f(\beta,\mu-U)
\bigr\} \nn\\
&\geq \bigl( \bar\alpha(\beta,\mu) - \bar\gamma(U) \bigr) |\partial \Lambda|, \nn
\end{align}
with
\ba
C_{d,\mu} &= \Bigl( \frac{4\pi\sqrt d}{|S_d|^{1/d}} + 2d (2d+1) \Bigr)
\frac1{1+\e{-\beta\mu}}, \nn\\
C_{d,\mu}' &= \frac{4\pi\sqrt d}{|S_d|^{1/d}}
\frac1{1+\e{-\beta(\mu-U)}}. \nn
\end{align}
\end{theorem}

The term $|\Omega|^{1-\frac1d}$ on the left side is not exactly proportional to
$|\partial\Lambda|$. However, we have in mind that $|\Lambda|$ and $|\Omega|$ are
comparable, in which case $|\Omega|^{1-\frac1d}$ is no greater than $|\partial\Lambda|$ (up
to a factor).

Notice that the upper bound vanishes as $\mu\to-\infty$, i.e.\ when the density tends to 0.
In the limit $U\to\infty$, Theorem \ref{thmpostemp} takes a simpler form, namely
\be
C_{d,\mu} |\partial\Lambda| \geq F_{\Omega,\Lambda}^{U=\infty}(\beta,\mu) - |\Lambda|
f(\beta,\mu) \geq \bar\alpha(\beta,\mu) |\partial\Lambda|.
\end{equation}

This extension of Theorem \ref{mainthm} to the case of positive (electronic) temperatures
is explained in Section \ref{secpostemp}. The lower bound follows from Propositions
\ref{proplbtemp} and \ref{proplbtempfinU}, while the upper bound is stated in Proposition
\ref{propubtemp}.

Our next step is to find upper and lower bounds for the total grand-canonical `free energy' by averaging over
$\Lambda$ (i.e., averaging over the positions of the classical particles). This can be done
with the aid of the Ising model free energy $f_{\rm Ising}(\beta, h)$,
\be
f_{\rm Ising}(\beta,h) = -\frac1\beta \lim_{|\Omega| \to \infty} \frac1{|\Omega|}
\sum_{\{s_x=\pm1\}} \exp\Bigl\{ -\beta \sumtwo{\{x,y\}\subset\Omega}{|x-y|=1} s_x s_y -
\beta h \sum_{x\in\Omega} s_x \Bigr\},
\end{equation}
where the sum is over configurations of classical spins on $\Omega$.

\begin{corollary}
\label{corIsing}
If $U$ is large enough (so that $\bar\alpha(\beta,\mu)-\bar\gamma(U)>0$), we have Ising bounds for the full free energy,
\bm
\tfrac12 [f(\beta,\mu) + f(\beta,\mu-U)] + \tfrac14 \bar\alpha + \tfrac{\bar\alpha}{4d} f_{\rm Ising}\bigl(
\tfrac1{4d} \bar\alpha \beta, \tfrac{2d}{\bar\alpha} [f(\beta,\mu) - f(\beta,\mu-U)] \bigr) \\
\leq -\frac1\beta \lim_{|\Omega| \to \infty} \frac1{|\Omega|} \log \sum_{\Lambda \subset
\Omega} \e{-\beta F_{\Omega,\Lambda}^U(\beta,\mu)} \\
\leq \tfrac12 [f(\beta,\mu) + f(\beta,\mu-U)] + \tfrac12 d C_{d,\mu} + \tfrac12 C_{d,\mu} f_{\rm Ising}\bigl(
\tfrac12 C_{d,\mu} \beta, \tfrac1{C_{d,\mu}} [f(\beta,\mu) - f(\beta,\mu-U)] \bigr) \nn
\end{multline}
where $\bar\alpha = \bar\alpha(\beta,\mu) - \bar\gamma(U)$.
\end{corollary}

The proof can be found at the end of Section \ref{secpostemp}.

Another consequence of Theorem \ref{thmpostemp} concerns the equilibrium state; namely, it
allows for a precise meaning of segregation. We consider the
probability that sites $x$ and $y$ are {\it both} occupied by classical particles, or both
are unoccupied. Namely, we consider
\be
\expval{\delta_{w_x,w_y}}_\Omega^{\phantom{x}} = \frac{\sum_{\Lambda \subset \Omega : w_x=w_y}
\exp \bigl\{ -\beta
F_{\Omega,\Lambda}^U(\beta,\mu) \bigr\}}{\sum_{\Lambda\subset\Omega} \exp \bigl\{ -\beta
F_{\Omega,\Lambda}^U(\beta,\mu) \bigr\}}
\end{equation}
where the sums are over subsets $\Lambda$ of $\Omega$ such that $|\Lambda| = [(1-n_{\rm c})
|\Omega|]$ ($[z]$ denotes the integer part of $z \in \bbR$). The restriction $w_x=w_y$
means that either both $x$ and $y$ belong to $\Lambda$, or both belong to the complement of
$\Lambda$.

Segregation means that up to a small fraction of sites that are close to the boundary
between classical particles and empty sites, any two sites at finite distance are either
both hosts of a classical particle, or are both empty. The fraction of sites close to the
boundary vanishes in the thermodynamic limit. Hence we {\it expect} that
\be
\lim_{\beta\to\infty} \lim_{|x-y|\to\infty} \lim_{|\Omega| \to \infty}
\expval{\delta_{w_x,w_y}}_\Omega^{\phantom{x}} = 1,
\end{equation}
but we are unable to prove it. 
Notice that using Theorem \ref{mainthm}, one can conclude that
\be
\label{weakresult}
\lim_{|x-y|\to\infty} \lim_{|\Omega| \to\infty} \lim_{\beta\to\infty}
\expval{\delta_{w_x,w_y}}_\Omega^{\phantom{x}} = 1.
\end{equation}
Indeed, taking the limit of zero temperature at finite volume, the sum over $\Lambda$
becomes restricted to the ground state configuration(s), whose boundary fraction
$|\partial\Lambda|/|\Lambda|$ tends to zero in the thermodynamic limit.

We can however take advantage of Theorem \ref{thmpostemp} to obtain a result that is better
than \eqref{weakresult}:

\begin{corollary}
\label{coreqst}
If $U$ is large enough (depending on $\mu$ and $d$ only), the ground state of the
Falicov-Kimball model displays segregation, in the sense that
$$
\lim_{|x-y|\to\infty} \lim_{\beta\to\infty} \lim_{|\Omega| \to\infty}
\expval{\delta_{w_x,w_y}}_\Omega^{\phantom{x}} = 1.
$$
\end{corollary}

The proof of this corollary can be found at the end of Section \ref{secpostemp}.

\section{The discrete Laplace operator in a finite domain}
\label{seclb}

\subsection{Basic properties}

We start our investigations by taking the limit $U\to\infty$. Let us denote $h_\Lambda \equiv
h_\Lambda^{U=\infty}$, the corresponding Hamiltonian, which acts on functions $\varphi \in
L^2(\Lambda)$ as follows: if $x \in \Lambda$,
\be
\label{defHaminfinU}
[h_\Lambda \varphi](x) = -\sum_{y \in \Lambda, |x-y|=1} \varphi(y) + 2d\,\varphi(x).
\end{equation}

Some observations can readily be made, that will be useful in the sequel. For
$\varphi \in L^2(\Lambda)$, one has the following formula,
\be
\label{gradient}
(\varphi, h_\Lambda \varphi) = \sum_{\{x,y\}: |x-y|=1} \bigl| \varphi(x) - \varphi(y)
\bigr|^2,
\end{equation}
where the sum is over {\it all} $x,y\in\bbZ^d$, with the understanding that
$\varphi(x) = 0$ for $x \notin \Lambda$. Equation \eqref{gradient} takes a simple
form because of the diagonal term in $h_\Lambda$. The effect of the Dirichlet
boundary conditions appears through a term $\sum_{x\in\partial\Lambda}
|\varphi(x)|^2$, that is due to pairs $\{x,y\}$ with $x\in\Lambda$ and
$y\notin\Lambda$.

The matrix $h_\Lambda$ is self-adjoint, and \eqref{gradient} shows that $h_\Lambda
\geq 0$. Its spectrum has a symmetry. Let $\varphi$ be an eigenvector with eigenvalue
$e$, and $\bar\varphi$ be the function $\bar\varphi(x) = (-1)^{|x|} \varphi(x)$; one
easily checks that $\bar\varphi$ is also an eigenvector, with eigenvalue $(4d-e)$.
The spectrum is therefore contained in the interval $[0,4d]$, and is symmetric
around $2d$. Furthermore, one has
\ba
&E_{\Lambda,|\Lambda|-N} = 2d \, |\Lambda| \, (1-2n) + E_{\Lambda,N}, \nn\\
&e(1-n) = 2d \, (1-2n) + e(n).
\end{align}
This allows to restrict ourselves to the case $n\leq \frac12$; indeed, existence of a lower
bound for a density $n$ implies a lower bound for the density $1-n$. This
symmetry holds only for $U=\infty$.

\subsection{The bulk term}

We are looking for a lower bound for the sum $E_{\Lambda,N}$ of the first $N$ eigenvalues of $h_\Lambda$.
This problem was considered by Li and Yau \cite{LY} for the Laplace operator in the
continuum. Let $\Lambda \subset \bbR^d$ be a bounded domain. They prove that the sum $S_N$ of
the first $N$ eigenvalues of the Laplace operator with Dirichlet boundary conditions is
bounded below,
\be
\label{LiYau}
S_N > (2\pi)^2 \frac d{d+2} |S_d|^{-\frac2d} N^{1+\frac2d} |\Lambda|^{-\frac2d},
\end{equation}
where $|S_d|$ and $|\Lambda|$ are the volumes of respectively the $d$-dimensional sphere and of $\Lambda$.

The corresponding inequality in the discrete case --- our Theorem \ref{mainthm} without the
boundary correction --- constitutes the heart of this paper, and
we explain below the proof of Li and Yau; see also \cite{LL}, Theorem 12.3.

It is useful to introduce the Fourier transform for functions in $L^2(\bbZ^d)$, that
is, on the whole lattice. A function $\varphi \in L^2(\Lambda)$ can be considered an
element of $L^2(\bbZ^d)$ by setting $\varphi(x)=0$ outside $\Lambda$. This Fourier
transform will lead to the electronic energy density for the infinite lattice, which
is the bulk term for the energy of electrons in $\Lambda$.

The Fourier transform of a function $\varphi$ is defined by
\be
\label{defFourtransf}
\hat\varphi(k) = \sum_{x \in \bbZ^d} \varphi(x) \e{\ii k x}, \quad k \in [-\pi,\pi]^d,
\end{equation}
and the inverse transform is
\be
\varphi(x) = \frac1{(2\pi)^d} \int_{[-\pi,\pi]^d} \dd k\, \hat\varphi(k) \e{-\ii k x}.
\end{equation}
Using the Fourier transform, a little thought shows that the energy of a particle in a state $\varphi$ in
$L^2(\Lambda)$ is
\be
(\varphi, h_\Lambda \, \varphi) = \frac1{(2\pi)^d}
\int_{[-\pi,\pi]^d} \dd k\, |\hat\varphi(k)|^2 \varepsilon_k,
\end{equation}
with $\varepsilon_k$ defined in \eqref{defeps} and with $\varphi(x) = 0$ if $x \notin
\Lambda$ in \eqref{defFourtransf}.

Let us consider $N$ orthonormal functions $\varphi_1, \dots, \varphi_N$, and let $E_\Lambda(\varphi_1, \dots,
\varphi_N)$ be their energy. We have
\be
\label{energie}
E_\Lambda(\varphi_1, \dots, \varphi_N) = \frac1{(2\pi)^d} \int_{[-\pi,\pi]^d} \dd k\, \rho(k)
\varepsilon_k,
\end{equation}
with
\be
\rho(k) = \sum_{j=1}^N |\hat\varphi_j(k)|^2.
\end{equation}

The function $\rho(k)$ satisfies the following equations:
\startalpheqno
\ba
\label{boundrho}
&0 \leq \rho(k) \leq |\Lambda|, \\
&\frac1{(2\pi)^d} \int_{[-\pi,\pi]^d} \dd k\, \rho(k) = N.
\end{align}
\stopalpheqno

\noindent
Indeed, positivity of $\rho$ is immediate and the last equation is Plancherel's identity. 
The upper bound \eqref{boundrho} for $\rho(k)$ can be seen by writing
\be
\rho(k) = (f, P \, f),
\end{equation}
where $P$ is the projector onto $\{\varphi_j\}_{j=1}^N$,
\be
P_{x,y} = \sum_{j=1}^N \varphi_j(x) \varphi_j^*(y),
\end{equation}
and $f$ is the vector
\be
f_x = \e{-\ii k x} \upchi_\Lambda(x).
\end{equation}
Then, since $P \leq \bbbone$, we have $\rho(k) \leq \|f\|^2 = |\Lambda|$.

Clearly, we have the lower bound
\be
\label{clearlb}
E_{\Lambda,N} \geq \inftwo{\rho:\; 0 \leq \rho \leq |\Lambda|}{(2\pi)^{-d} \int \rho = N} \frac1{(2\pi)^d} \int_{[-\pi,\pi]^d}
\dd k\, \rho(k) \, \varepsilon_k.
\end{equation}
We can use the bathtub principle (\cite{LL}, Theorem 1.14) to find the infimum: it is given by the function
\be
\label{rhomin}
\rho_{\rm min}(k) = \begin{cases} |\Lambda| & \text{if } \varepsilon_k \leq \varepsilon_{\text F} \\ 0 & \text{otherwise,} \end{cases}
\end{equation}
where the Fermi level $\varepsilon_{\text F}$ is given by the relation
$\frac1{(2\pi)^d} \int_{\varepsilon_k < \varepsilon_{\text F}} \dd k = N/|\Lambda|$.
Thus the right side of \eqref{clearlb} is precisely equal to $|\Lambda| \, e(N/|\Lambda|)$.

\section{Lower bound involving the boundary}
\label{seclbb}

In the previous section, we showed that $E_{\Lambda,N}$ is bounded below by its bulk term. Now we
strengthen this inequality and prove that $E_{\Lambda,N}$ also includes a term proportional to the
boundary of $\Lambda$. This can be checked for $d=1$ by explicit computation, but higher dimensions require more elaborate treatment.

We start with a lemma that applies when the density $n$ is small enough (or, by the
symmetry for $h_\Lambda$, when it is close to 1).

\begin{lemma}
\label{lemlbld}
If $n \leq |S_d|/(4\pi)^d$, we have
$$
E_{\Lambda,N} \geq |\Lambda| e(n) + \frac{2^{d-3} n^{1+\frac2d}}{\pi^d d^3 |S_d|^{2/d}}
|\partial\Lambda|.
$$
\end{lemma}

\begin{proof}
Recall that
\be
E_{\Lambda,N} = \frac1{(2\pi)^d} \int_{[-\pi,\pi]^d} \dd k \, \rho(k) \varepsilon_k
\end{equation}
with $\rho(k) = \sum_{j=1}^N |\hat\varphi(k)|^2$. We want to show that $\rho(k)$ cannot be
too close to $\rho_{\rm min}(k)$ in \eqref{rhomin}. By completeness of the set of
eigenvectors $\{\varphi_j\}$, we have
\be
\rho(k) = |\Lambda| - \sum_{j=N+1}^{|\Lambda|} |\hat\varphi(k)|^2.
\end{equation}

We now use the Schr\"odinger equation. We have $h_\Lambda \varphi_j(x) = e_j \varphi_j(x)$ for $x\in\Lambda$. Let
us take $\varphi_j(x)=0$ for $x\notin\Lambda$; then the following equation holds
true for all $x\in\bbZ^d$:
\be
-\sum_e \varphi_j(x+e) + \upchi_{\Lambda^\compl}(x) \sum_{e: x+e \in \Lambda} \varphi_j(x+e) + 2d
\varphi_j(x) = e_j \varphi_j(x).
\label{Schreq}
\end{equation}
The middle term of the left side is necessary for the equation to hold at sites outside
$\Lambda$, that are neighbors of sites in $\Lambda$.  Taking the Fourier transform, we get
\be
\varepsilon_k \hat\varphi_j(k) + (b_k,\varphi_j) = e_j \hat\varphi_j(k),
\label{SchreqFour}
\end{equation}
where $b_k$ is a `boundary vector',
\be
b_k(x) = \upchi_{\partial\Lambda}(x) \e{-\ii kx} \sum_{e: x+e \notin \Lambda} \e{-\ii ke}.
\label{defboundvec}
\end{equation}
Notice that $|\partial \Lambda| \leq \|b_k\|^2 \leq (2d)^2 |\partial \Lambda|$ if
$|k|_\infty \leq \frac\pi3$. From \eqref{SchreqFour}, we have
\be
|\hat\varphi_j(k)|^2 = \frac{|(b_k,\varphi_j)|^2}{(\varepsilon_k-e_j)^2} \geq \frac1{(4d)^2}
|(b_k,\varphi_j)|^2.
\label{ineqphiS}
\end{equation}

The electronic energy in $\Lambda$ is given by $\int\rho(k)\varepsilon_k$. We saw in
Section \ref{seclb} that $0 \leq \rho(k) \leq |\Lambda|$. By \eqref{defboundvec} and
\eqref{ineqphiS}, this can be strengthened to
\be
\const \, \|P_- b_k\|^2 \leq \rho(k) \leq |\Lambda| - \const \, \|P_+ b_k\|^2,
\end{equation}
where $P_-$ (resp.\ $P_+$) is the projector onto the subspace spanned by $(\varphi_1, \dots,
\varphi_N)$ (resp.\ $(\varphi_{N+1}, \dots,
\varphi_{|\Lambda|})$). See Fig.\ \ref{figbathtub} for intuition. We show below that
\be
\| P_+ b_k \|^2 \geq \const \, |\partial\Lambda|,
\end{equation}
and this will straightforwardly lead to the lower bound.

\bfig
\epsfxsize=100mm
\centerline{\epsffile{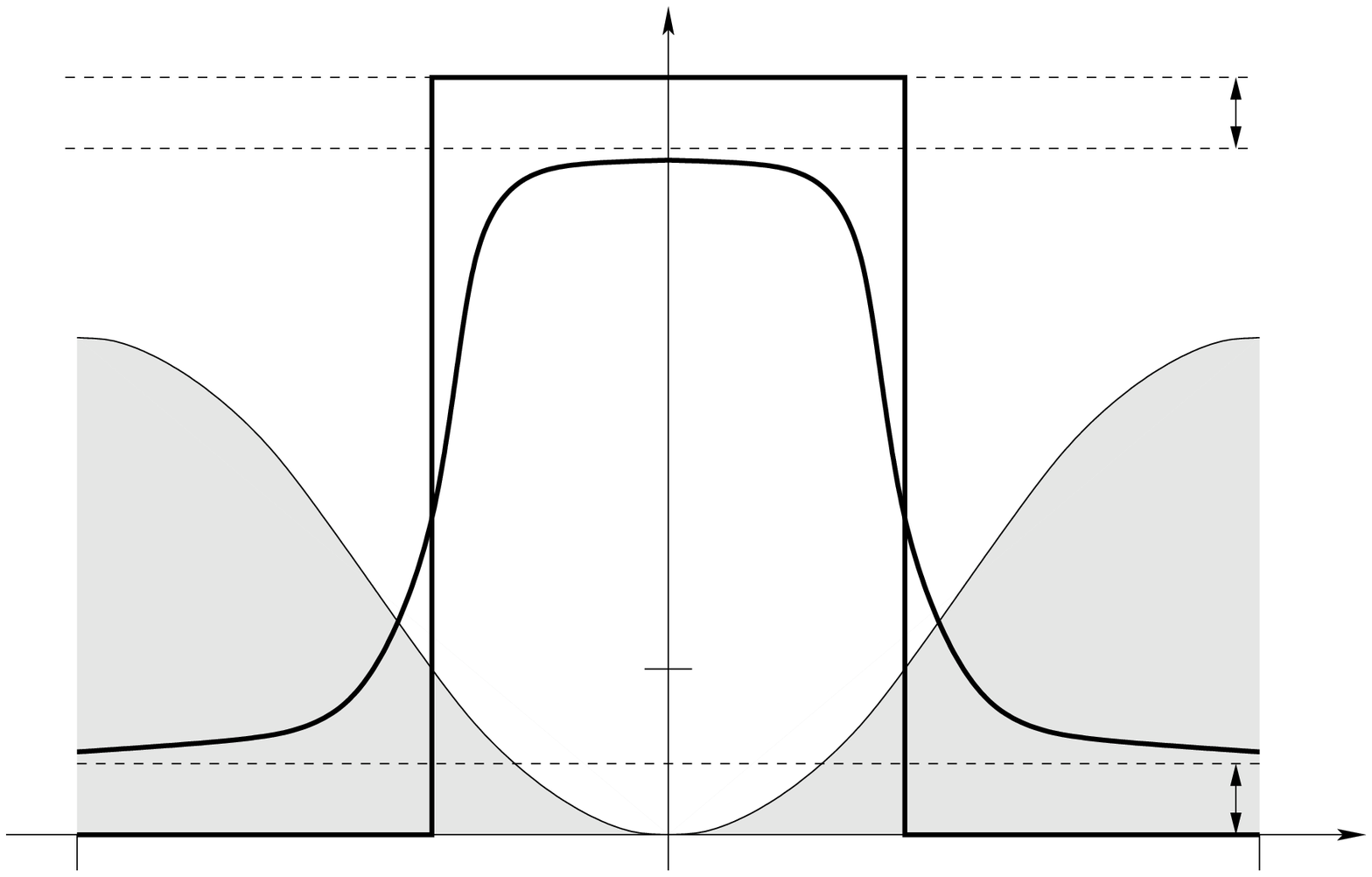}}
\figtext{
\writefig	-4.6	0.4	{$-\pi$}
\writefig	-0.2	0.4	{\footnotesize 0}
\writefig	3.93	0.4	{$\pi$}
\writefig	4.7	0.5	{$k$}
\writefig	-0.6	5.97	{\footnotesize $|\Lambda|$}
\writefig	1.7	5.3	{\small $\rho_{\rm min}(k)$}
\writefig	0.5	4.7	{\small $\rho(k)$}
\writefig	4.2	4.35	{$\varepsilon_k$}
\writefig	-0.7	2.05	{\small $\varepsilon_{\rm F}$}
\writefig	4.1	1.1	{\footnotesize $\const \, \|P_- b_k \|^2$}
\writefig	4.1	5.9	{\footnotesize $\const \, \|P_+ b_k \|^2$}
}
\caption{Illustration of the expression \eqref{energie} for $E_{\Lambda,N}$;
$\rho(k)$ satisfies more stringent estimates than those
stated in \eqref{boundrho}, and this plays an important role in
deriving the lower bound.}
\label{figbathtub}
\end{figure}

In order to see that the boundary vector has a projection in the subspace of the
eigenvectors with large eigenvalues, we first remark that
\be
(b_k, h_\Lambda b_k) = \sum_{\{x,y\}: |x-y|=1} |b_k(x) - b_k(y)|^2 \geq \| b_k \|^2.
\end{equation}
We used the fact that each site $x$ of $\partial\Lambda$ has at least one neighbor $y$
outside of $\Lambda$, and we obtained an inequality by restricting the sum over such pairs.
Let us introduce $N'$ such that $e_{N'} \leq \frac12$ and $e_{N'+1} >
\frac12$. We first consider the situation where $N \leq N'$. Using first $e_j \leq 4d$ and then the previous inequality, we have
\ba
4d \sum_{j= N'+1}^{|\Lambda|} |(b_k,\varphi_j)|^2 &\geq \sum_{j= N'+1}^{|\Lambda|}
|(b_k,\varphi_j)|^2 e_j \nn\\
&\geq \|b_k\|^2 - \sum_{j=1}^{N'} |(b_k,\varphi_j)|^2 e_j \geq
\tfrac12 \|b_k\|^2.
\label{one_more_bound}
\end{align}
For $|k|_\infty \leq \frac\pi3$, \eqref{ineqphiS} and \eqref{one_more_bound} imply
\be
\label{encore_une_borne}
\rho(k) \leq |\Lambda| - \frac{|\partial\Lambda|}{2(4d)^3}.
\end{equation}

We can write a lower bound by proceeding as in Section \ref{seclb}, but using the 
bound (\ref{encore_une_borne})
for $\rho(k)$, instead of $|\Lambda|$. The bathtub principle then gives
\be
\label{bathbound}
E_{\Lambda,N} - |\Lambda| e(n) \geq \frac1{(2\pi)^d} \int_{\varepsilon_{\rm F} <
\varepsilon_k < \varepsilon_{\rm F}'} \dd k \, \varepsilon_k \Bigl( |\Lambda| -
\frac{|\partial\Lambda|}{2(4d)^3} \Bigr) - \frac1{(2\pi)^d} \int_{\varepsilon_k <
\varepsilon_{\rm F}} \dd k \, \varepsilon_k \frac{|\partial\Lambda|}{2(4d)^3},
\end{equation}
where we introduce $\varepsilon_{\rm F}'$ such that
\be
N = \frac1{(2\pi)^d} \Bigl( |\Lambda| - \frac{|\partial\Lambda|}{2(4d)^3} \Bigr)
\int_{\varepsilon_k < \varepsilon_{\rm F}'} \dd k.
\end{equation}
Notice that for $n \leq |S_d| / (4\pi)^d$, we have $\varepsilon_{\rm F}' < \frac12$, so that
$\varepsilon_k < \varepsilon_{\rm F}'$ implies $|k|_\infty < \frac\pi3$. This justifies the
use of \eqref{encore_une_borne}.
We bound the first integral of \eqref{bathbound} using $\varepsilon_k > \varepsilon_{\rm
F}$, and we obtain
\be
E_{\Lambda,N} - |\Lambda| e(n) \geq \frac{|\partial\Lambda|}{2(4d)^3} \frac1{(2\pi)^d} \int_{\varepsilon_k < \varepsilon_{\rm
F}} \dd k (\varepsilon_{\rm F} - \varepsilon_k).
\end{equation}

One can derive a more explicit expression for the lower bound. First,
\be
\int_{\varepsilon_k < \varepsilon_{\rm F}} \dd k ( \varepsilon_{\rm F} - \varepsilon_k) \geq
\tfrac12 \varepsilon_{\rm F} \int_{\varepsilon_k \leq \frac12 \varepsilon_{\rm F}} \dd k.
\end{equation}
Second we use $1-\frac{\theta^2}2 \leq \cos\theta \leq 1 - \frac4{\pi^2} \theta^2$, to get
\be
\label{boundsepsk}
\frac8{\pi^2} |k|^2 \leq \varepsilon_k \leq |k|^2.
\end{equation}
One can use the upper bound of \eqref{boundsepsk} to get
\be
\int_{\varepsilon_k < \frac12 \varepsilon_{\rm F}} \dd k \geq |S_d| (\tfrac12
\varepsilon_{\rm F})^{d/2}.
\end{equation}
Recall that $|S_d|$ is the volume of the unit sphere in $d$ dimensions.
The lower bound of \eqref{boundsepsk} allows to write
\be
\varepsilon_{\rm F} \geq \frac{2^5 n^{2/d}}{|S_d|^{2/d}}.
\end{equation}
Then one gets the bound
\be
\int_{\varepsilon_k < \varepsilon_{\rm F}} \dd k ( \varepsilon_{\rm F} - \varepsilon_k) \geq
\frac{2^{4+2d} n^{1+\frac2d}}{|S_d|^{2/d}}.
\end{equation}
Hence the boundary correction to $E_{\Lambda,N}$ is bounded below by $\alpha(n)
|\partial\Lambda|$ with
\be
\alpha(n) = \frac{2^{d-3}}{\pi^d d^3 |S_d|^{2/d}} n^{1+\frac2d}.
\end{equation}

Recall that we supposed $N \leq N'$, where $N'$ is the index of the largest
eigenvalue that is smaller than $\frac12$. Were it not the case, we can write, with $n'=N'/|\Lambda|$,
\be
E_{\Lambda,N} = \sum_{j=1}^{N'} e_j + \sum_{j=N'+1}^N e_j \geq |\Lambda|
e(n') + \alpha(n') |\partial\Lambda| + \tfrac12 |\Lambda| (n-n').
\end{equation}
We used the previous inequality to bound the first sum, and $e_j \geq \frac12$ for the
second sum. This is greater than $|\Lambda| e(n) + \alpha(n) |\partial\Lambda|$ provided
\be
e(n') + \alpha(n') \frac{|\partial\Lambda|}{|\Lambda|} + \tfrac12 (n-n')
\geq e(n) + \alpha(n) \frac{|\partial\Lambda|}{|\Lambda|}.
\end{equation}
A sufficient condition is that $\frac12 n - e(n) - \alpha(n)$ is an increasing function of
$n$. Computing the derivative (the derivative of $e(n)$ is $\varepsilon_{\rm F}(n)$, that
is smaller than $(2\pi)^2 n^{2/d}/|S_d|^{2/d}$ using \eqref{boundsepsk}), and requiring it
to be positive leads to the condition
\be
n \leq \frac{|S_d|}{\{ 2(2\pi)^2 + \frac{2^{d-2}}{\pi^d d^3} (1+\frac2d) \}^{d/2}}.
\end{equation}
The right side is greater than $|S_d|/(4\pi)^d$.
\end{proof}

It may seem obvious that the extra energy due to the presence of the boundary increases
as $n$ increases, until it reaches $\frac12$. But we can provide no proof for this, and
hence we need a new derivation for the lower bound with higher densities. We proceed in two
steps. First we give a lemma that works when the boundary has few nearest neighbors; the
proof is similar to that of the previous lemma. Then we give three lemmas, with more
intricate demonstrations, and that establish the lower bound for boundaries where at least
a density of sites have nearest neighbors. We need some notation to characterize the
configuration around a site $x$ of the boundary.

Let $e,e'$ be unit vectors in $\bbZ^d$; the notation $e \parallel i$ means that $e$ is
parallel to the $i$-th direction; equivalently, the
components of $e$ are given by $e_k = \pm
\delta_{ik}$. We introduce integers $q_{x,i}$ and $q_{x,ij}$; for $x \in
\partial \Lambda$, we set
\ba
q_{x,i} &= \#\{ e \parallel i : x+e \notin \Lambda \} \nn\\
q_{x,ij} &= \#\{ (e,e'): e \parallel i, e' \parallel j, \, x+e \in \partial \Lambda, \, x+e+e' \notin
\Lambda \}.
\end{align}
Notice that $0 \leq q_{x,i} \leq 2$ and $0 \leq q_{x,ij} \leq 4$. Also, $q_{x,ii} = \#\{ e
\parallel i, x+e \in \partial \Lambda, x+2e \notin \Lambda \}$, and $0 \leq q_{x,ii} \leq
2$. We also define $q_x = \sum_i q_{x,i}$.

The following lemma applies to domains where
most boundary sites $x$ satisfy $q_{x,ij} \equiv 0$, in which case $x$ has no neighbors
that belong to the boundary. Here, $q_{x,ij} \equiv 0$ means that, at $x$, $q_{x,ij}=0$
for all $1\leq i,j\leq d$.

\begin{lemma}
\label{lemdiagby}
For all $\Lambda \subset \bbZ^d$ with
$$
\# \{ x \in \partial \Lambda : q_{x,ij} \not\equiv 0 \} \leq \frac1{32 d^4} |\partial \Lambda|,
$$
there exists $\alpha(n)>0$ such that
$$
E_{\Lambda,N} \geq |\Lambda| e(n) + \alpha(n) |\partial \Lambda|.
$$
\end{lemma}

Remark: $\lim_{n\to0} \alpha(n) = 0$ and $\alpha(1-n) = \alpha(n)$ by the symmetry for
$h_\Lambda$.

\begin{proof}
We can suppose $N \leq \frac{|\Lambda|}2$. The definition of $\rho(k)$ involves a sum over
the first $N$ eigenvectors (more precisely, of their Fourier transforms). In case of
degenerate eigenvalues one is free to choose any eigenvectors. For the proof of Lemma
\ref{lemdiagby} it turns out that the possible degeneracy of $2d$ brings some burden, and it
is useful to redefine $\rho(k)$ by averaging over eigenvectors with eigenvalue $2d$:
\be
\rho(k) = \begin{cases} \sum_{j=1}^N |\hat\varphi_j(k)|^2 & \text{if } N \leq \tilde N \\
\sum_{j=1}^{\tilde N} |\hat\varphi_j(k)|^2 + \frac{N-\tilde N}{|\Lambda|-2\tilde N} \sum_{j:
e_j=2d} |\hat\varphi_j(k)|^2 & \text{if } N > \tilde N; \end{cases}
\end{equation}
here, $\tilde N$ is such that $e_{\tilde N} < 2d$ and $e_{\tilde N+1} \geq 2d$. The
degeneracy of $2d$ is $|\Lambda| - 2 \tilde N$ (which may be zero). Of course,
$E_{\Lambda,N}$ is still given as the integral of $\rho(k)$ multiplied by $\varepsilon_k$.

The goal is to prove that $\rho(k)$ cannot approach $\rho_{\rm min}$ in \eqref{rhomin}.
Since $\sum_{j=1}^{|\Lambda|} |\hat\varphi_j(k)|^2 = |\Lambda|$, we have
\be
\rho(k) \leq |\Lambda| - \Bigl\{ \sum_{j = |\Lambda|-\tilde N+1}^{|\Lambda|} |\hat\varphi_j(k)|^2 + \frac12 \sum_{j:
e_j=2d} |\hat\varphi_j(k)|^2 \Bigr\}.
\label{eqrho}
\end{equation}

We introduce
\be
S(k) = \sum_{j=|\Lambda|-\tilde N+1}^{|\Lambda|} |(b_k,\varphi_j)|^2 + \frac12 \sum_{j:
e_j=2d} |(b_k,\varphi_j)|^2,
\end{equation}
with $b_k$ the boundary vector defined in \eqref{defboundvec}.
By the inequality \eqref{ineqphiS}, it is enough to show that $S(k)$ is bounded below by a
quantity of the order of $|\partial\Lambda|$. We have
\be
S(k) = \bigl( b_k, P_+ \,b_k \bigr) + \tfrac12 (b_k, P_0 \, b_k),
\end{equation}
where $P_+$
is the projector onto the subspace spanned by all $\varphi_j$ with $e_j>2d$, and $P_0$
is the projector corresponding to the eigenvalue $2d$.

We want to show that $S(k) \geq \const |\partial \Lambda|$ for small $|k|$. This amounts to
prove that the vector $b_k$ cannot lie entirely in the subspace spanned by $\{ \varphi_j
\}_{1\leq j\leq N}$.

Notice that if $x \in \partial \Lambda$ and $q_{x,ij} \equiv 0$, then $x$ has no neighbors in
$\partial \Lambda$. Using the assumption of Lemma \ref{lemdiagby}, as well as $|b_k(x)| \leq 2d$
and $|\partial \Lambda| \leq \| b_k \|^2$, we get
\ba
\label{B1}
(b_k, h_\Lambda \, b_k) &= \sum_{\{x,y\}: |x-y|=1} |b_k(x)-b_k(y)|^2 \nn\\
&\geq 2d \sum_{x \in \partial \Lambda} |b_k(x)|^2 - 2d \sumtwo{x \in \partial \Lambda}{q_{x,ij}
\not\equiv 0} |b_k(x)|^2 \\
&\geq \bigl( 2d - \frac1{4d} \bigr) \|b_k\|^2.\nn
\end{align}
The last inequality uses the assumption of Lemma \ref{lemdiagby}, and the fact that
$|b_k(x)|$ is at most $2d$ and at least 1.

Next we consider $\|(h_\Lambda-2d) b_k\|^2$. We have, for $x\in \Lambda$,
\be
\bigl[ (h_\Lambda - 2d) b_k \bigr](x) = -\sum_e b_k(x+e),
\end{equation}
and therefore, if $|k|_\infty \leq \frac\pi3$,
\be
\label{B2}
\|(h_\Lambda-2d) b_k\|^2 \geq \sum_{x \in \Lambda} \sum_e |b_k(x+e)|^2 = \sum_x (2d-q_x) |b_k(x)|^2.
\end{equation}

We write $b_k=b'+b''$, with $b''(x) = b_k(x)$ if $q_x=2d$, 0 otherwise. Notice that $b'\perp
b''$. Clearly, $P_0 b'' = b''$, and therefore
\be
S(k) = (b', P_+ \,b') + \tfrac12 (b', P_0 \, b') + \tfrac12 \|b''\|^2.
\label{bGb}
\end{equation}
Furthermore, from \eqref{B1} and \eqref{B2}, $b'$ satisfies
\be
\label{B3}
(b', (h_\Lambda - 2d) b') \geq -\frac1{4d} \|b_k\|^2, \quad\quad
(b', (h_\Lambda-2d)^2 b') \geq \|b'\|^2.
\end{equation}

Because $|e_j-2d| \leq 2d$, the last inequality implies
\be
-\sum_j |(\varphi_j,b')|^2 (e_j-2d) + 2\sum_{j: e_j>2d} |(\varphi_j,b')|^2 (e_j-2d) = \sum_j |(\varphi_j,b')|^2 |e_j-2d| \geq 
\frac{\|b'\|^2}{2d}.
\end{equation}
With the first inequality in \eqref{B3}, this yields
\be
\sum_{j: e_j>2d} |(\varphi_j,b')|^2 (e_j-2d) \geq \frac{\|b'\|^2}{4d} - \frac{\|b_k\|^2}{8d},
\end{equation}
hence
\be
\sum_{j: e_j>2d} |(\varphi_j,b')|^2 \geq \frac{\|b'\|^2}{8d^2} -\frac{\|b_k\|^2}{16d^2}.
\end{equation}
Back to \eqref{bGb}, we obtain
\be
S(k) \geq \frac{\|b'\|^2}{8d^2} + \tfrac12 \|b''\|^2 -\frac{\|b_k\|^2}{16d^2} \geq
\frac{\|b_k\|^2}{16d^2}.
\end{equation}

We can combine this bound with \eqref{eqrho} and \eqref{ineqphiS}; we have then for all
$|k|_\infty < \frac\pi3$
\be
\rho(k) \leq |\Lambda| - \frac{|\partial \Lambda|}{(4d)^4}.
\end{equation}

We introduce $\tilde \varepsilon_{\text F}$ such that
\be
\frac1{(2\pi)^d} \Bigl( |\Lambda| - \frac{|\partial \Lambda|}{(4d)^4} \Bigr) \int_{\varepsilon_k <
\tilde \varepsilon_{\text F},|k|_\infty < \frac\pi3}
\dd k +
\frac{|\Lambda|}{(2\pi)^d} \int_{\varepsilon_k <
\tilde \varepsilon_{\text F},|k|_\infty > \frac\pi3} \dd k = N,
\end{equation}
and we have
\bm
E_{\Lambda,N} - |\Lambda| e(\tfrac N{|\Lambda|}) \geq \frac1{(2\pi)^d} \Bigl( |\Lambda| -
\frac{|\partial \Lambda|}{(4d)^4} \Bigr) \int_{\varepsilon_{\text F} < \varepsilon_k
< \tilde \varepsilon_{\text F},|k|_\infty < \frac\pi3} \dd k \, \varepsilon_k \\
+ \frac{|\Lambda|}{(2\pi)^d} \int_{\varepsilon_{\text F} < \varepsilon_k
< \tilde \varepsilon_{\text F},|k|_\infty > \frac\pi3} \dd k \, \varepsilon_k
- \frac1{(2\pi)^d} \frac{|\partial \Lambda|}{(4d)^4} \int_{\varepsilon_k
< \varepsilon_{\text F},|k|_\infty < \frac\pi3} \dd k \, \varepsilon_k.
\end{multline}
We bound the first two integrals using $\varepsilon_k > \varepsilon_{\text F}$; from the
definitions of $\varepsilon_{\text F}$ and $\tilde \varepsilon_{\text F}$ we have
\be
\frac{|\partial\Lambda|}{(2\pi)^d} \int_{\varepsilon_k < \tilde \varepsilon_{\rm F}, |k|_\infty <
\frac\pi3} \dd k = \frac{|\Lambda|}{(2\pi)^d} \int_{\varepsilon_{\rm F} < \varepsilon_k <
\tilde \varepsilon_{\rm F}} \dd k.
\end{equation}
As a result, we obtain the bound we were looking for,
\be
E_{\Lambda,N} - |\Lambda| e(\tfrac N{|\Lambda|}) \geq \frac{|\partial \Lambda|}{(4d)^4} \frac1{(2\pi)^d}
\int_{\varepsilon_k < \varepsilon_{\text F}, |k|_\infty < \frac\pi3} \dd k \, \bigl(
\varepsilon_{\text F} - \varepsilon_k \bigr).
\end{equation}
\end{proof}

We present now another lemma that claims the lower bound for $E_{\Lambda,N}$, and that involves a
new assumption. We shall see below in Lemmas \ref{lemboundfluct} and \ref{lemmu} that for all volumes, at least
one of these lemmas applies.

\begin{lemma}
\label{lemfluct}
Let $\delta>0$ and $n \geq |S_d|/(4\pi)^d$. We assume that
$$
\| (h_\Lambda-e_N) b_{k_0} \|^2 \geq \delta |\partial \Lambda|,
$$
for {\rm some} $k_0$ belonging to the Fermi surface, i.e.\ $\varepsilon_{k_0} = \varepsilon_{\rm
F}$ where $\varepsilon_{\rm F}$ is the Fermi energy for density $n=\frac N{|\Lambda|}$.
Then we have
$$
E_{\Lambda,N} - |\Lambda| e(n) \geq \eta \, |\partial \Lambda|
$$
with $\eta = |S_d|^5 \delta^{30d+2} / (2^{271d+23} \pi^{10d+2} d^{130d+9})$.
\end{lemma}

The constant $\eta$ that appears as a lower bound seems ridiculously small, but we cannot
do better. Notice that this lower bound is much smaller than the one obtained in Lemma
\ref{lemlbld} at low density, with $n = |S_d|/(4\pi)^2$. We expect however that the lower
bound is an increasing function of $n$ for $0\leq n\leq\frac12$, although we cannot prove it.

\begin{proof}
We have
\be
E_{\Lambda,N} - |\Lambda| e(n) = \frac1{(2\pi)^d} \int_{[-\pi,\pi]^d} \dd k \Bigl[ \Delta_-(k)
\varepsilon_k - \Delta_+(k) \varepsilon_k \Bigr],
\end{equation}
where
\ba
\Delta_+(k) &= \bigl( |\Lambda| - \rho(k) \bigr) \charfct{\varepsilon_k <
\varepsilon_{\rm F}} \nn\\
\Delta_-(k) &= \rho(k) \charfct{\varepsilon_k > \varepsilon_{\rm F}}. \nn
\end{align}
We used here another convention for the characteristic function, namely
$\charfct{\bf\cdot}$ is 1 if ${\bf\cdot}$ is true, and is 0 otherwise.
Notice that $\int\dd k \,\Delta_-(k) = \int\dd k \,\Delta_+(k)$. Then we both have
\be
E_{\Lambda,N} - |\Lambda| e(n) \geq \begin{cases} \frac1{(2\pi)^d} \int\dd k \, \Delta_-(k)
(\varepsilon_k - \varepsilon_{\rm F}) \\ \frac1{(2\pi)^d} \int\dd k \, \Delta_+(k)
(\varepsilon_{\rm F} - \varepsilon_k). \end{cases}
\end{equation}
And by H\"older, this implies
\be
\label{lowbd}
E_{\Lambda,N} - |\Lambda| e(n) \geq \Bigl( \frac1{(2\pi)^d} \int\dd k \, [\Delta_\pm(k)]^{1/5}
\Bigr)^5 \Big/ \Bigl( \frac1{(2\pi)^d} \int\dd k \, |\varepsilon_k - \varepsilon_{\rm
F}|^{-\frac14} \Bigr)^4.
\end{equation}
One shows in Lemma \ref{lemapp} (a) that the integral of $|\varepsilon_k - \varepsilon_{\rm
F}|^{-\frac14}$ is bounded by 2.

Recall that $\{\varphi_j\}_{1
\leq j \leq |\Lambda|}$ are the eigenvectors of $h_\Lambda$. Let $P_-$, resp.\ $P_+$, be the projectors
onto the first $N$ eigenvectors, resp.\ the last $|\Lambda|-N$ eigenvectors. By \eqref{ineqphiS},
one has inequalities
\ba
\Delta_+(k) &\geq \frac1{(4d)^2} \| P_+ b_k \|^2 \quad\quad \text{if } \varepsilon_k <
\varepsilon_{\text F}, \nn\\
\Delta_-(k) &\geq \frac1{(4d)^2} \| P_- b_k \|^2 \quad\quad \text{if } \varepsilon_k >
\varepsilon_{\text F}.
\label{delta}
\end{align}

Let us introduce sets $A$ and $A'$ by
\ba
A &= \{ k: \varepsilon_k < \varepsilon_{\rm F} \text{ and } |k-k_0| <
\tfrac{\delta^3}{2^{25} d^{25/2}} \}, \nn\\
A' &= \{ k: \varepsilon_k > \varepsilon_{\rm F} \text{ and } |k-k_0| <
\tfrac{\delta^3}{2^{25} d^{25/2}} \}.
\end{align}
We obtain a lower bound by substituting \eqref{delta} into \eqref{lowbd}, and restricting
the integrals to $A$ and $A'$. Namely,
\ba
E_{\Lambda,N} - |\Lambda| e(n) &\geq \frac1{2^8 d^2} \Bigl( \frac1{(2\pi)^d} \int_A \dd k
\, \|P_+ b_k\|^{2/5} \Bigr)^5, \nn\\
E_{\Lambda,N} - |\Lambda| e(n) &\geq \frac1{2^8 d^2} \Bigl( \frac1{(2\pi)^d} \int_{A'} \dd k
\, \|P_- b_k\|^{2/5} \Bigr)^5. \nn
\end{align}

Let $\tilde b_k = b_k / \|b_k\|$. From the assumption of the lemma, and using $\|
h_\Lambda-\varepsilon \| \leq 4d$ and Lemma
\ref{lemapp} (d), we have that for all $k \in A\cup A'$,
\be
\label{boinf}
\frac{\|b_k\|^2}{|\partial\Lambda|} \geq \frac\delta{2^5 d^2}.
\end{equation}
Extracting a factor $|\partial\Lambda|$, and using the above inequality, we can write
\ba
E_{\Lambda,N} - |\Lambda| e(n) &\geq |\partial\Lambda| \frac\delta{2^{13} d^4 (2\pi)^{5d}} \Bigl( \int_A \dd k
\, \|P_+ \tilde b_k\|^{2/5} \Bigr)^5, \nn\\
E_{\Lambda,N} - |\Lambda| e(n) &\geq |\partial\Lambda| \frac\delta{2^{13} d^4 (2\pi)^{5d}} \Bigl( \int_{A'} \dd k
\, \|P_- \tilde b_k\|^{2/5} \Bigr)^5.
\label{bornes}
\end{align}

Consider $k\in A$. The assumption of the lemma for $k_0$, together with the bound for the
gradient in Lemma \ref{lemapp} (e), implies
\be
\frac{(b_k, (h_\Lambda-e_N)^2 b_k)}{|\partial\Lambda|} \geq \frac\delta2.
\end{equation}
Therefore
\be
(\tilde b_k, (h_\Lambda-e_N)^2 \tilde b_k) \geq \frac\delta{8d^2}.
\end{equation}
This can be rewritten as
\be
\sum_{j=1}^{|\Lambda|} |(\varphi_j, \tilde b_k)|^2 (e_j-e_N)^2 \geq \frac\delta{8d^2},
\end{equation}
that is,
\be
\sum_{j=1}^{|\Lambda|} |(\varphi_j, \tilde b_k)|^2 (e_j^2+e_N^2) \geq \frac\delta{8d^2}
+ 2e_N (\tilde b_k, h_\Lambda \, \tilde b_k).
\end{equation}
Hence
\be
(\tilde b_k, h_\Lambda \, \tilde b_k) \leq e_N + \sum_{j=1}^{|\Lambda|} |(\varphi_j, \tilde
b_k)|^2 \bigl( \frac{e_j^2}{2e_N} - \frac{e_N}2 \bigr) - \frac\delta{2^5 d^3}.
\end{equation}
The quantity in the brackets is negative for $j\leq N$. Observing that $e_N \geq
\frac{e(|S_d|/(4\pi)^d)}{|S_d|/(4\pi)^d} \geq 1 / 2^{d+1} \pi^2$ (because $n
\geq |S_d|/(4\pi)^d$ and using
Lemma \ref{lemapp} (c)), the bracket is bounded by $2^{d+4} \pi^2 d^2$.
Therefore,
\be
(\tilde b_k, h_\Lambda \, \tilde b_k) \leq e_N + 2^{d+4} \pi^2 d^2 \, \|P_+
\tilde b_k\|^2 - \frac\delta{2^5 d^3}.
\end{equation}

On the other hand, for $k' \in A'$,
\be
(\tilde b_{k'}, h_\Lambda \, \tilde b_{k'}) \geq \sum_{j=N+1}^{|\Lambda|} |(\varphi_j, \tilde b_{k'})|^2 e_j \geq
e_{N+1} - 4d \|P_- \tilde b_{k'}\|^2.
\end{equation}
Since $|k-k'| \leq \delta^3/2^{24} d^{25/2}$, we have from Lemma \ref{lemapp} (g) and
\eqref{boinf}
\be
(\tilde b_{k'}, h_\Lambda \, \tilde b_{k'}) - (\tilde b_k, h_\Lambda \, \tilde b_k) \leq
\frac\delta{2^6 d^3}.
\end{equation}
Therefore
\be
e_{N+1} - 4d \|P_- \tilde b_{k'}\|^2 \leq
e_N + 2^{d+4} \pi^2 d^2 \|P_+ \tilde b_k\|^2 - \frac\delta{2^5 d^3} + \frac\delta{2^6 d^3}.
\end{equation}
Clearly, $e_N \leq e_{N+1}$; then
\be
\|P_+ \tilde b_k\|^2 + \tfrac1{2^{d+2} \pi^2 d} \|P_- \tilde b_{k'}\|^2 \geq
\frac\delta{2^{d+10} \pi^2 d^5}.
\end{equation}

We use now \eqref{bornes}. The worst situation happens when $\|P_+ \tilde b_k\|^2$ is
equal to the right side of the previous equation. Using Lemma \ref{lemapp} (b) we finally
get the lower bound of Lemma \ref{lemfluct}.
\end{proof}

Now we show that we can use Lemma \ref{lemfluct}
for all $\Lambda$ such that Lemma \ref{lemdiagby} does not apply.

Let $a_x = \bigl( (2d-\varepsilon) q_{x,i} \bigr)_{1\leq i\leq d}$ and $Q_x =
\bigl( (1+\delta_{ij}) q_{x,ij} \bigr)_{1\leq i,j\leq d}$. More generally, we let $a$ denote a vector with entries
$(2d-\varepsilon) q_{x,i}$, and $Q$ a matrix with entries $2q_{x,ii}$ in the diagonal and
$q_{x,ij}$ off the diagonal, that correspond to a
possible configuration around $x$. With $c = (\cos k_i)_{1\leq i\leq d}$, we introduce
\be
F(c;a,Q) = (a,c) + \tfrac12 \Tr Q - (c,Qc).
\end{equation}
This function appears when establishing a lower bound for $\| (h_\Lambda-\varepsilon) b_k
\|^2$.

Let $\caQ$ be the set of all matrices $Q$ (for which there exists some compatible
configuration); we introduce
\be
\caQ'' = \{ Q \in \caQ: Q_{ii} \equiv 2 \text{ and } Q_{ij} + Q_{ji} = 4 \text{ for all }
i\neq j \}
\end{equation}
and
\be
\caQ' = \{ Q \notin \caQ'': Q_{ij} \not\equiv 0 \}.
\end{equation}
The reason behind the definition of $\caQ'$ is that we can provide a lower bound only if
$F(c;a,Q)$ is not uniformly zero when $k$ moves along the Fermi surface (i.e.\ with
$\varepsilon_k = \varepsilon_{\rm F}$); and we can show that $F(c;a,Q)$ is not uniformly
zero only for $Q \in \caQ'$, see Lemma \ref{lemmu} below.

For given $\varepsilon_{\rm F}$, we define
\be
\mu(\varepsilon_{\rm F}) = \min_{a,Q\in\caQ'} \min_{\varepsilon \in [0,2d]}
\max_{c: \varepsilon_k = \varepsilon_{\rm F}} |F(c;a,Q)|.
\end{equation}
We state a lower bound
involving $\mu(\varepsilon_{\rm F})$, and check below in Lemma \ref{lemmu} that
$\mu(\varepsilon_{\rm F})$ is strictly positive for $\varepsilon_{\rm F} > 0$.

\begin{lemma}
\label{lemboundfluct}
Let $d\geq2$. For all finite $\Lambda$ satisfying
$$
\#\{ x \in \partial \Lambda : q_{x,ij} \not\equiv 0 \} \geq \frac1{32d^4} |\partial
\Lambda|,
$$
we have
$$
\max_{k: \varepsilon_k = \varepsilon_{\rm F}} \| (h_\Lambda-\varepsilon) b_k \|^2 \geq
\frac{\mu(\varepsilon_{\rm F})}{2^6 d^5 5^{d^2}} |\partial \Lambda|.
$$
\end{lemma}

The factor $\frac1{32 d^4}$ is arbitrary here, and has been taken such in order to
complement the condition of Lemma \ref{lemdiagby}.

\begin{proof}
Let us introduce
\be
q_k(x) = \upchi_{\partial\Lambda}(x) \sum_{e: x+e \notin \Lambda} \e{-\ii ke}.
\end{equation}
By the definition of the discrete Laplacian,
\be
\bigl[ (h_\Lambda - \varepsilon) b_k \bigr](x) = \e{-\ii kx} \Bigl\{ (2d-\varepsilon) q_k(x) -
\sum_e \e{-\ii ke} q_k(x+e) \Bigr\}.
\end{equation}
Let us denote by $r_k(x)$ the quantity inside the brackets above. Clearly, $\|(h_\Lambda-\varepsilon)
b_k\|^2 = \|r_k\|^2$. Let $\caR_a$, $a = 1,\dots, 2^d$, represents all combinations of inversions of some
coordinates. We have the following inequality:
\be
\label{lowerbound}
\frac1{2^d} \sum_{a=1}^{2^d} \|r_{\caR_a k}\|^2 \geq \Bigl\| \frac1{2^d}
\sum_{a=1}^{2^d} r_{\caR_a k} \Bigr\|^2.
\end{equation}
Indeed, starting from the RHS, we have in essence (with $0\leq a_i \leq 1$ and $\sum_i a_i
= 1$)
\ba
\bigl( \sum_i a_i \vec v_i, \sum_i a_i \vec v_i \bigr) &= \sum_{i,j} a_i a_j (\vec v_i, \vec
v_j) \nn\\
&\leq \bigl( \sum_i \sqrt{a_i} \sqrt{a_i} \|\vec v_i\| \bigr)^2 \\
&\leq \Bigl[ \Bigl( \sum_i a_i \Bigr)^{1/2} \Bigl( \sum_i a_i \|\vec v_i\|^2 \Bigr)^{1/2}
\Bigr]^2 \nn
\end{align}
which is the LHS of \eqref{lowerbound}.

The RHS of \eqref{lowerbound} is clearly smaller than $\max_{k: \varepsilon_k =
\varepsilon_{\rm F}} \| (h_\Lambda-\varepsilon) b_k \|^2$.
One computes now $\sum_{a=1}^{2^d} r_{\caR_a k}(x)$ for $x \in \partial \Lambda$. First,
\be
\label{someeq}
\frac1{2^d} \sum_{a=1}^{2^d} (2d-\varepsilon) q_{\caR_a k}(x) = (2d-\varepsilon)
\sum_{i=1}^d q_{x,i} \cos k_i.
\end{equation}
Second,
\ba
\label{uneeq}
-\frac1{2^d} \sum_{a=1}^{2^d} \sum_{e : x+e \in \partial \Lambda} &\e{-\ii \caR_a k \,e} \sum_{e'
: x+e+e' \notin \Lambda} \e{-\ii \caR_a k \,e'} \nn\\
&= -\sum_{i=1}^d q_{x,ii} \cos(2k_i) -
\frac12 \sum_{i,j: i \neq j} q_{x,ij} \bigl[ \cos(k_i+k_j) + \cos(k_i-k_j) \bigr] \\
&= -2\sum_{i=1}^d q_{x,ii} \cos^2 k_i + \sum_{i=1}^d q_{x,ii} -
\sum_{i,j: i \neq j} q_{x,ij} \cos k_i \cos k_j. \nn
\end{align}
We used $\cos(2k_i) = 2\cos^2 k_i - 1$, and the bracket in the second line is $[\cdot] = 2\cos k_i \cos
k_j$. 

Gathering \eqref{someeq} and \eqref{uneeq}, we obtain
\be
\frac1{2^d} \sum_{a=1}^{2^d} r_{\caR_a k}(x) =  F(c;a_x,Q_x).
\end{equation}

One can check that whenever $Q_x \notin \caQ'$ and differs from 0, there exists a neighbor
$y$ that belongs to $\caQ'$. Then the condition of the lemma implies that
\be
\#\{ x \in \partial\Lambda: Q_x \in \caQ' \} \geq \frac1{2^6 d^5} |\partial\Lambda|.
\end{equation}
Furthermore, $\caQ'$ has less than $5^{d^2}$ elements since $0\leq Q_{ij}\leq 4$; then for
any $\Lambda$ that satisfies the assumption of the lemma there exists $Q \in \caQ'$ such
that
\be
\#\{ x \in \partial\Lambda: Q_x=Q \} \geq \frac1{2^6 d^5 5^{d^2}} |\partial\Lambda|.
\end{equation}
We get a lower bound for $\| (h_\Lambda-\varepsilon) b_k \|^2$ by considering only those
sites, i.e.\
\be
\max_{k: \varepsilon_k = \varepsilon_{\rm F}} \|(h_\Lambda-\varepsilon) b_k\|^2 \geq
\sum_{x\in\partial\Lambda: Q_x=Q} \max_{c: \varepsilon_k = \varepsilon_{\rm F}}
|F(c;a_x,Q_x)| \geq \frac{\mu(\varepsilon_{\rm F})}{2^6 d^5 5^{d^2}} |\partial\Lambda|
\end{equation}
uniformly in $\varepsilon \in [0,2d]$.
\end{proof}

There remains to be checked that $\mu(\varepsilon_{\rm F})$ differs from 0.

\begin{lemma}
\label{lemmu}
For all $\varepsilon_{\rm F}>0$, we have $\mu(\varepsilon_{\rm F}) \neq 0$.
\end{lemma}

\begin{proof}
We proceed {\it ab absurdo} and explore ways where
$F(c;a,Q)$ could be uniformly zero.

The constraint $\varepsilon_k = \varepsilon_{\rm F}$ takes a simple form, namely $(u,c) = d-\frac12
\varepsilon_{\text F}$. Furthermore, $c$ satisfies $|c|_\infty \leq 1$; if
$\varepsilon_{\rm F} \neq 0$, we can find $\delta c$ such that $|c+\delta c|_\infty \leq 1$
and $(u,c+\delta c) = d - \frac12 \varepsilon_{\rm F}$ --- in which case $\delta c$
must be perpendicular to $u$.
The condition $F(c + \delta c; a,Q) = F(c;a,Q)$ for all $\delta c \perp u$ implies that
$a-2Qc \parallel u$. This should also be true when $c$ is replaced with $c + \delta c$,
hence $Q \delta c \parallel u$ for all $\delta c \perp u$. Now take $(\delta c)_\ell =
\delta_{i\ell} - \delta_{j\ell}$. We have
\ba
(Q \delta c)_i &= Q_{ii} - Q_{ij} \nn\\
(Q \delta c)_j &= -Q_{jj} + Q_{ji},
\end{align}
and these two components must be equal, since $Q \delta c$ is parallel to $u$. Hence
$Q_{ii} + Q_{jj} = Q_{ij}+Q_{ji}$, or
\be
\label{condforvanish}
2q_{x,ii} + 2q_{x,jj} = q_{x,ij} + q_{x,ji}.
\end{equation}
In this case $F(c;a,Q)$ takes the form
\be
F(c;a,Q) = (2d-\varepsilon) \sum_{i=1}^d q_{x,i} c_i - (2d - \varepsilon_{\rm F}) \sum_{i=1}^d
q_{x,ii} c_i + \sum_{i=1}^d q_{x,ii}.
\end{equation}

Since $Q\in\caQ'$ we have $q_{x,ij} \not\equiv 0$; if $(u,c)=0$, one can take $c=0$, and
$F(c;a,Q)$ is strictly positive, so we can suppose $(u,c)\neq0$.

Let $s = \sum_i q_{x,ii}/(u,c)$, and $v$ the vector with components
\be
v_i = (2d-\varepsilon) q_{x,i} - (2d-\varepsilon_{\rm F}) q_{x,ii} + s.
\end{equation}
Then $F(c;a,Q) = (v,c)$. If we require this to be zero for $c \parallel u$, then we need $v
\perp u$. But we also require $(v,c+\delta c) = (v,c)$ for all $\delta c \perp u$, hence $v
\parallel u$. So $v$ must be zero, i.e.\
\be
(2d-\varepsilon) q_{x,i} - (2d-\varepsilon_{\rm F}) q_{x,ii} + s = 0
\end{equation}
for all $1\leq i\leq d$.

We also have $q_{x,i} + q_{x,ii} \leq 2$, and $q_{x,ii}$ cannot be always
equal to 2. If $s\neq0$, one checks that necessarily $q_{x,ii} \equiv 1$, which is
impossible because $Q\in\caQ'$. Hence $F(c;a,Q)$ cannot be uniformly zero when moving along
the Fermi surface.
\end{proof}

\section{Finite $U$}
\label{secfinU}

We consider now the Falicov-Kimball model with finite repulsion $U$, and establish a lower
bound for the ground state energy of $N$ electrons in a configuration specified by
$\Lambda$. More precisely, we show that when decreasing the repulsion $U$, one does not
lower the energy more than $\const \cdot |\partial\Lambda|/U$.

For any $\Lambda$, the spectrum of $h_{\Omega,\Lambda}^U$ is included in $[0,4d] \cup
[U,U+4d]$. When $U>4d$, eigenstates with energy in $[0,4d]$ show exponential decay outside
of $\Lambda$; and eigenstates with energy in $[U,U+4d]$ show exponential decay inside
$\Lambda$. Hence $\Lambda$ and $\Omega\setminus\Lambda$ are essentially decorrelated, and
the situation is close to that with $U=\infty$.

The following proposition compares
the energies of electrons with finite and infinite $U$. It is useful to introduce
$\eta(U)$,
\be
\label{defeta}
\eta(U) = \Bigl( \frac{2d}{U-2d} \Bigr)^2 \sum_{j=1}^d
\Bigr[ \frac{(U-2d)^2}{U(U-4d)} \Bigr]^j = \frac{(U-2d)^{2d}}{(U(U-4d))^d} - 1.
\end{equation}
Notice that $\lim_{U\to\infty} U^2 \eta(U) = 4d^3$, as it easily comes out from the middle
expression.

\begin{proposition}
\label{propfinU}
If $U>4d$, we have
$$
E_{\Lambda,N}^U \geq E_{\Lambda,N} - \gamma(U) |\partial \Lambda|,
$$
with
$$
\gamma(U) = \frac{8d^2}{U-2d} + d \, 2^{d+2} \eta(U).
$$
\end{proposition}

\begin{proof}
First, we remark that eigenvectors of $h_{\Omega,\Lambda}^U$ with eigenvalue smaller than
$4d$ have
exponential decay outside of $\Lambda$. Indeed, for $x \notin \Lambda$ the Schr\"odinger equation can be written
\be
\varphi_j(x) = \frac{\sum_e \varphi_j(x+e)}{U+2d-e_j}.
\end{equation}
If $e_j \leq 4d$, we have
\be
\sum_{j=1}^{|\Lambda|} |\varphi_j(x)|^2 \leq \frac{\sum_{j=1}^{|\Lambda|} 2d \sum_e
|\varphi_j(x+e)|^2}{(U-2d)^2}.
\end{equation}

Using this inequality, we can proceed by induction on the distance between $x$ and $\Lambda$. The
induction hypothesis is that the following holds true
\be
\sum_{j=1}^{|\Lambda|} |\varphi_j(x)|^2 \leq \Bigl( \frac{2d}{U-2d} \Bigr)^{2n}
\end{equation}
for any $x$ such that $\dist(x,\Lambda) \geq n$. As a result, we have
\be
\label{expdecay}
\sum_{j=1}^{|\Lambda|} |\varphi_j(x)|^2 \leq \Bigl( \frac{2d}{U-2d} \Bigr)^{2 \, \dist(x,\Lambda)}.
\end{equation}

Let us introduce
\be
\tilde\rho_{xy} = \sum_{j=1}^N \upchi_\Lambda(x) \, \varphi_j^*(x) \, \varphi_j(y) \,
\upchi_\Lambda(y).
\end{equation}
We show that $E^U_{\Lambda,N}$ is bounded below by $\Tr \tilde\rho h_\Lambda$, up to a
contribution no greater than $\const \, |\partial\Lambda|/U$. Recall that $h_\Lambda$ is the
Hamiltonian with infinite repulsions. If $P_\Lambda$ is the projector onto the
domain $\Lambda$, let $\tilde\varphi_j = P_\Lambda \varphi_j$.
\ba
E^U_{\Lambda,N} &= \sum_{j=1}^N \Bigl( \sum_{\{x,y\} : |x-y|=1} |\varphi_j(x) - \varphi_j(y)|^2 + U
\sum_{x \notin \Lambda} |\varphi_j(x)|^2 \Bigl) \nn\\
&\geq \sum_{j=1}^N \Bigl( \sum_{\{x,y\}: |x-y|=1} |\tilde\varphi_j(x) - \tilde\varphi_j(y)|^2
+ \sumtwo{\{x,y\} \not\subset \Lambda}{|x-y|=1} |\varphi_j(x) - \varphi_j(y)|^2 -
\sumtwo{x \in \Lambda, y \notin \Lambda}{|x-y|=1} |\varphi_j(x)|^2 \Bigl) \nn\\
&\geq \Tr \tilde\rho h_\Lambda - 2 \sum_{j=1}^N \sumtwo{x\in\Lambda,
y\notin\Lambda}{|x-y|=1} |\varphi_j(x)| \, |\varphi_j(y)|.
\label{borninf}
\end{align}
By the Schwarz inequality, the last term is smaller than
\be
\label{B6}
2 \Bigl( \sumtwo{x\in\Lambda, y\notin\Lambda}{|x-y|=1} \sum_{j=1}^N |\varphi_j(x)|^2
\Bigr)^{1/2} \Bigl( \sumtwo{x\in\Lambda, y\notin\Lambda}{|x-y|=1} \sum_{j=1}^N |\varphi_j(y)|^2
\Bigr)^{1/2} \leq \frac{8d^2}{U-2d} |\partial\Lambda|.
\end{equation}
We used \eqref{expdecay} with $\dist(x,\Lambda)$ being respectively 0 and 1, in order to
control the quantities in both brackets.

Recall that $e_j$ denotes the $j$-th eigenvalue of the Hamiltonian $h_\Lambda$; that is, with
infinite repulsions. Let us introduce the projector $P_j$ onto the corresponding
eigenstate. Then
\be
\Tr \tilde\rho h_\Lambda = \sum_{j=1}^{|\Lambda|} e_j \, \Tr \tilde\rho P_j \equiv \sum_{j=1}^{|\Lambda|} e_j \, n_j,
\end{equation}
where the $n_j$ satisfy $0 \leq n_j \leq 1$, and $\sum_j n_j = \Tr \tilde\rho$. By the
bathtub principle \cite{LL}, we obtain the lower bound
\be
\label{B4}
\Tr \tilde\rho h_\Lambda \geq \sum_{j=1}^{\Tr \tilde\rho} e_j.
\end{equation}

There remains to show that $\Tr \tilde\rho$ is close to $N$. We have
\ba
N - \Tr \tilde\rho &= \sum_{x \notin \Lambda} \sum_{j=1}^N |\varphi_j(x)|^2 \leq \sum_{n=1}^\infty
\#\{ x: \dist(x,\Lambda)=n \} \Bigl( \frac{2d}{U-2d} \Bigr)^{2n} \nn\\
&\leq |\partial \Lambda|
\sum_{n=1}^\infty 2^d \bigl( \begin{smallmatrix} n+d-1 \\ d-1 \end{smallmatrix} \bigr)
\Bigl( \frac{2d}{U-2d} \Bigr)^{2n} \label{someupperbound} = 2^d \eta(U) |\partial\Lambda|.
\end{align}
We bounded $\#\{\cdot\} \leq 2^d \bigl( \begin{smallmatrix} n+d-1 \\ d-1 \end{smallmatrix}
\bigr) |\partial \Lambda|$. Since $e_j \leq 4d$, we obtain the proposition.
\end{proof}

\section{Upper bound}
\label{secub}

We establish now an upper bound for the sum of the first $N$ eigenvalues in a finite domain
$\Lambda$, for the case of infinite repulsion. The bound carries over to finite $U$, since
$E_{\Lambda,N}^U$ is increasing in $U$.

The strategy is to average $h_\Lambda$ over a huge box. The `strength' of the averaged
Hamiltonian depends on the number of bonds in $\Lambda$, which is roughly
$2d|\Lambda|-|\partial\Lambda|$. The averaged Hamiltonian is, up to a factor, the hopping matrix
in the huge box, and its ground state energy is easy to compute in the thermodynamic
limit. This can be compared to $E_{\Lambda,N}$ by concavity of the sum of lowest
eigenvalues of self-adjoint operators. The result is

\begin{proposition}
\label{propub}
The sum of the first $N$ eigenvalues of the Laplace operator in a domain $\Lambda$ with
Dirichlet boundary conditions, satisfies the upper bound
$$
E_{\Lambda,N} \leq |\Lambda| e(n) + |\partial\Lambda| (2dn-e(n)).
$$
\end{proposition}

\begin{proof}
Let $L$ be a multiple of $|\Lambda|$, and $N_L$ be such that $N_L/L^d = N/|\Lambda|$. We consider a box
$\{1,\dots,L\}^d$. We introduce $\tilde\varepsilon = \frac12 (e_N + e_{N+1})$. Let
$\caR_a$, $a=1,\dots,L^d d!$, represent a translation possibly followed by an axis
permutation. We define the averaged Hamiltonian
\be
\bar h_{L,\Lambda} = \frac1{L^d d!} \sum_{a=1}^{L^d d!} \Bigl( h_{\caR_a \Lambda} -
\tilde\varepsilon \bbbone_{\caR_a \Lambda} \Bigr).
\end{equation}
Then
\ba
S_{N_L}(\bar h_{L,\Lambda}) &\geq \frac1{L^d d!} \sum_{a=1}^{L^d d!} S_{N_L} \bigl( h_{\caR_a \Lambda} -
\tilde\varepsilon \bbbone_{\caR_a \Lambda} \bigr) \nn\\
&= S_N(h_\Lambda - \tilde\varepsilon \bbbone_\Lambda).
\label{boundconcav}
\end{align}
Indeed, all summands in the above equation are equal, and the Hamiltonian $h_\Lambda -
\tilde\varepsilon \bbbone_\Lambda$ has no more than $N$ negative eigenvalues, and at least
$L^d-|\Lambda|$ zero eigenvalues. The RHS is equal to $E_{\Lambda,N} - N \tilde\varepsilon$.

Let $K_i$ be the number of sites in $\Lambda$ that have
$i$ neighbors in $\Lambda$. We have $|\Lambda|=\sum_{i=0}^{2d} K_i$ and $|\partial \Lambda| =
\sum_{i=0}^{2d-1} K_i$; and the number of bonds in $\Lambda$
is $\frac12 \sum_{i=0}^{2d} i K_i$. Then the averaged Hamiltonian is
\be
\bigl( \bar h_{L,\Lambda} \bigr)_{xy} = -\frac t{L^d} \delta_{|x-y|=1} +
(2d-\tilde\varepsilon) \frac{|\Lambda|}{L^d} \delta_{xy},
\end{equation}
with
\be
t = \frac1{2d} \sum_{i=0}^{2d} i K_i.
\end{equation}

Let $K = \sum_{i=0}^{2d} \frac{2d-i}{2d} K_i$; then $t = |\Lambda|-K$ and $K \leq
|\partial\Lambda|$. One easily checks that
\be
\bar h_{L,\Lambda} = \frac{|\Lambda|}{L^d} h_{\{1,\dots,L\}^d} + \frac K{L^d} \bigl( 2d \bbbone_{\{1,\dots,L\}^d} -
h_{\{1,\dots,L\}^d} \bigr) - \tilde\varepsilon \frac{|\Lambda|}{L^d} \bbbone_{\{1,\dots,L\}^d}.
\end{equation}
Notice that all operators commute. In \eqref{boundconcav}, the terms involving
$\tilde\varepsilon$ cancel, since $S_{N_L}(\bbbone_{\{1,\dots,L\}^d}) = N_L$, and $N_L
\frac{|\Lambda|}{L^d} = N$. Now, as $L \to \infty$,
\be
\frac1{L^d} S_{N_L}(h_{\{1,\dots,L\}^d}) \to e(n).
\end{equation}
Therefore \eqref{boundconcav} implies
\be
|\Lambda| e(n) + K (2dn - e(n)) \geq E_{\Lambda,N}.
\end{equation}
\end{proof}

\section{Positive electronic temperature}
\label{secpostemp}

This section considers the electronic free energy at positive temperature, for a fixed
configuration of classical particles. We will see that the inequalities satisfied by the
sums over lowest eigenvalues have an extension to free energies.

\subsection{Lower bound for $U=\infty$}

We start with $U\to\infty$. Let $F_\Lambda(\beta,\mu) = -\frac1\beta \log\Tr \e{-\beta
H_\Lambda}$ where the trace is taken in the Fock space of antisymmetric wave functions on
$\Lambda$, and $H_\Lambda \equiv H_{\Omega,\Lambda}^{U=\infty}$ is the second
quantized form of the one-particle Hamiltonian $h_\Lambda$ defined by \eqref{defHaminfinU}.

\begin{proposition}
\label{proplbtemp}
For all finite $\Lambda$, one has the lower bound
$$
F_\Lambda(\beta,\mu) - |\Lambda| f(\beta,\mu) \geq \bar\alpha(\beta,\mu) |\partial\Lambda|,
$$
where $\bar\alpha(\beta,\mu) > 0$ satisfies $\lim_{\beta\to\infty} \bar\alpha(\beta,\mu) > 0$ if $0<\mu<4d$.
\end{proposition}

\begin{proof}
The fermionic free energy $F_\Lambda(\beta,\mu)$ can be expressed in terms of the
eigenvalues of $h_\Lambda$,
\be
F_\Lambda(\beta,\mu) = -\frac1\beta \sum_{j=1}^{|\Lambda|} \log\bigl( 1 + \e{-\beta (e_j-\mu)} \bigr).
\end{equation}
In order to compare this with the corresponding infinite-volume expression
\eqref{thermofen}, we partition the Brillouin zone $[-\pi,\pi]^d$ according to the level sets of the function $\varepsilon_k$;
more precisely, we define measures $\mu_j$, $1\leq j\leq |\Lambda|$, by
\be
\dd\mu_j(k) = \frac{|\Lambda|}{(2\pi)^d} \; \charfct{\varepsilon_{\rm F}(\tfrac{j-1}{|\Lambda|}) <
\varepsilon(k) < \varepsilon_{\rm F}(\tfrac j{|\Lambda|})} \dd k.
\end{equation}
Notice that $\int\dd\mu_j(k) = 1$ and $\frac1{|\Lambda|} \sum_{j=1}^{|\Lambda|} \dd\mu_j(k) = \frac{\dd
k}{(2\pi)^d}$.
Next we introduce $e_j^*$, that are equal to $\varepsilon_k$ averaged over $\mu_j$:
\be
e_j^* = \int\dd\mu_j(k) \, \varepsilon_k.
\label{defestar}
\end{equation}
The ground state energy \eqref{defbulken} of a density $N/|\Lambda|$ of electrons in $\bbZ^d$ can then be written as
\be
e(N/|\Lambda|) = \frac1{|\Lambda|} \sum_{j=1}^N e_j^*.
\end{equation}
 From the lower bound without a boundary term, we have
\be
\sum_{j=1}^N e_j > \sum_{j=1}^N e_j^*,
\label{discrineq}
\end{equation}
for all $N < |\Lambda|$, and equality when $N=|\Lambda|$.

Actually, inequality \eqref{discrineq} can be strengthened by introducing a term depending
on the boundary of $\Lambda$. In Theorem \ref{mainthm}, $\alpha(n)$ can be taken to be increasing in $n$ for $n\leq\frac12$. Also,
$\alpha(1-n) = \alpha(n)$. Therefore there exists a function $a(\varepsilon)$, with
$a(\varepsilon)>0$ for $0<\varepsilon<2d$, $a(4d-\varepsilon)=-a(\varepsilon)$, and
\be
\alpha(n) = \frac1{(2\pi)^d} \int_{\varepsilon_k < \varepsilon_{\rm F}} \dd k\,
a(\varepsilon_k).
\end{equation}

Next we define
\be
e_j' = \int\dd\mu_j(k) \, \bigl( \varepsilon_k + \tfrac{|\partial\Lambda|}{|\Lambda|} a(\varepsilon_k)
\bigr);
\end{equation}
then the following is stronger than \eqref{discrineq} and holds true,
\be
\sum_{j=1}^N e_j \geq \sum_{j=1}^N e_j'.
\end{equation}
With $a(\varepsilon)$ chosen appropriately
both sequences $(e_j)$ and $(e_j')$ are increasing, and the inequality above is an equality
when $N=|\Lambda|$. The sequence $(e_j)$ is said to `majorize' $(e_j')$.
We can apply an inequality due to Hardy, Littlewood and
P\'olya (and independently found by Karamata); see \cite{Mit} page 164.
For any concave function $g$, we have
\be
\label{ineq}
\sum_{j=1}^{|\Lambda|} g(e_j) \geq \sum_{j=1}^{|\Lambda|} g(e_j').
\end{equation}
(Conversely, if \eqref{ineq} holds for all concave $g$, then $(e_j)$ majorizes $(e_j')$.) We use this
inequality with
\be
g(e) = -\frac1\beta \log(1 + \e{-\beta(e-\mu)}),
\end{equation}
which is concave. We get
\be
F_\Lambda(\beta,\mu) \geq \sum_{j=1}^{|\Lambda|} g(e_j') \geq \frac{|\Lambda|}{(2\pi)^d}
\int_{[-\pi,\pi]^d} \dd k \, g \bigl( \varepsilon_k + \tfrac{|\partial\Lambda|}{|\Lambda|}
a(\varepsilon_k) \bigr),
\end{equation}
where the last step is Jensen's inequality. Then
\be
\frac1{|\Lambda|} F_\Lambda(\beta,\mu) - f(\beta,\mu) \geq \frac1{(2\pi)^d} \int \dd k \, \bigl\{
g \bigl( \varepsilon_k + \tfrac{|\partial\Lambda|}{|\Lambda|} a(\varepsilon_k) \bigr) - g(\varepsilon_k) \bigr\}.
\end{equation}

In the limit $\beta\to\infty$, we have
\be
g(e) = \begin{cases} e-\mu & \text{if } e<\mu \\ 0 & \text{if } e \geq \mu. \end{cases}
\end{equation}
As a result, for all $0<\mu<4d$ we get a lower bound for large $\beta$ that is uniform in
the limit $\beta\to\infty$.

One also gets a lower bound by using concavity of $g$, that holds for all temperatures, but
that is not uniform in $\beta$:
\ba
\frac1{|\Lambda|} F_\Lambda(\beta,\mu) - f(\beta,\mu) &\geq \frac1{(2\pi)^d} \int \dd k
\int_{\varepsilon_k}^{\varepsilon_k + \tfrac{|\partial\Lambda|}{|\Lambda|} a(\varepsilon_k)} \dd e \, g'(e) \nn\\
&= \frac1{(2\pi)^d} \int \dd k \Bigl\{ \tfrac{|\partial\Lambda|}{|\Lambda|} a(\varepsilon_k)
g'(\varepsilon_k) - O(a(\varepsilon_k)^2)
\Bigr\} \nn\\
&= \frac{|\partial\Lambda|}{|\Lambda|} \frac1{(2\pi)^d} \int_{\varepsilon_k < 2d} \dd k \,
a(\varepsilon_k) \Bigl\{ g'(\varepsilon_k) -
g'(4d-\varepsilon_k) - O(a(\varepsilon_k)) \Bigr\}.
\end{align}
The integrand in the last line is strictly positive if $a(\varepsilon_k)$ is
small enough, and chosen to vanish appropriately as $\varepsilon_k \to 2d$.
\end{proof}

\subsection{Lower bound with finite $U$}

We extend now the results of the previous section to the case of finite repulsion $U$. As
we noted in Section \ref{secfinU}, when $U>4d$ all eigenstates have exponential decay,
either in $\Lambda$ or in $\Omega\setminus\Lambda$. We show that the total free energy in
$\Omega$ is equal to a term involving $\Omega\setminus\Lambda$ only, plus a term involving
$\Lambda$ only, up to a correction of order $|\partial\Lambda|/U$.

\begin{proposition}
\label{proplbtempfinU}
$$
F_{\Omega,\Lambda}^U(\beta,\mu) \geq F_\Lambda(\beta,\mu) +
F_{\Omega\setminus\Lambda}(\beta,\mu-U) - \bar\gamma(U) |\partial\Lambda|
$$
with
$$
\bar\gamma(U) = (2dU + 4d + 8d^2) 2^d \eta(U) + \frac{(4d)^2}{U-2d},
$$
and $\eta(U)$ is defined in \eqref{defeta}.
Notice that $\lim_{U\to\infty} U \bar\gamma(U) = 16d^2 + 8d^4 2^d$.
\end{proposition}

\begin{proof}
Let us introduce
\be
\tilde\varphi_j(x) = \begin{cases} \varphi_j(x) & \text{if } 1\leq j\leq |\Lambda| \text{ and }
x\in\Lambda, \text{ or if } |\Lambda|<j\leq |\Omega| \text{ and } x \notin\Lambda \\ 0 &
\text{otherwise.} \end{cases}
\end{equation}
We assume $N>|\Lambda|$ (otherwise, replace $|\Lambda|$ by $N$ in the next expressions, and ignore the sums
whose initial number is greater than the final one).
Then
\ba
\sum_{j=1}^N e_j^U &= \sum_{j=1}^{|\Lambda|} \Bigl( \sumtwo{\{x,y\}}{|x-y|=1} |\tilde\varphi_j(x) -
\tilde\varphi_j(y)|^2 + \sumtwo{\{x,y\} \not\subset \Lambda}{|x-y|=1} |\varphi_j(x) -
\varphi_j(y)|^2 \nn\\
&\hspace{5cm} - \sumtwo{x\in\Lambda, y\notin\Lambda}{|x-y|=1} |\varphi_j(x)|^2 + U
\sum_{x\notin\Lambda} |\varphi_j(x)|^2 \Bigr) \nn\\
&+ \sum_{j=|\Lambda|+1}^N \Bigl( \sumtwo{\{x,y\}}{|x-y|=1} |\tilde\varphi_j(x) -
\tilde\varphi_j(y)|^2 + \sumtwo{\{x,y\} \not\subset \Lambda^\compl}{|x-y|=1} |\varphi_j(x) -
\varphi_j(y)|^2 \\
&\hspace{5cm} - \sumtwo{x\notin\Lambda, y\in\Lambda}{|x-y|=1} |\varphi_j(x)|^2 + U
\sum_{x\notin\Lambda} |\varphi_j(x)|^2 \Bigr). \nn
\end{align}
We proceed as in Section \ref{secfinU} and define
\ba
\tilde\rho_{xy} &= \sum_{j=1}^{|\Lambda|} \upchi_\Lambda(x) \varphi_j^*(x) \varphi_j(y)
\upchi_\Lambda(y), \nn\\
\tilde\rho_{xy}' &= \sum_{j=|\Lambda|+1}^N \upchi_{\Lambda^\compl}(x) \varphi_j^*(x) \varphi_j(y)
\upchi_{\Lambda^\compl}(y).
\end{align}
Then
\be
\sum_{j=1}^N e_j^U \geq \Tr\tilde\rho h_\Lambda + \Tr\tilde\rho'
[h_{\Omega\setminus\Lambda} + U] - 2 \sum_{j=1}^N \sumtwo{x\in\Lambda,
y\notin\Lambda}{|x-y|=1} |\varphi_j(x)| \, |\varphi_j(y)|.
\end{equation}
The inequality \eqref{B4} is still valid, for both $\tilde\rho$ and $\tilde\rho'$. Hence
\be
\label{B5}
\sum_{j=1}^N e_j^U \geq \sum_{j=1}^{\Tr\tilde\rho} e_j + \sum_{j=|\Lambda|+1}^{|\Lambda|+1+\Tr\tilde\rho'} (\bar e_j
+ U) - 2 \sum_{j=1}^N \sumtwo{x\in\Lambda, y\notin\Lambda}{|x-y|=1} |\varphi_j(x)| \,
|\varphi_j(y)|.
\end{equation}
Here, $\bar e_j$, $|\Lambda|<j\leq|\Omega|$ are the eigenvalues of the operator
$h_{\Omega\setminus\Lambda}$.

We define
\be
\delta_j = \begin{cases} 4d \sum_{x \notin \Lambda} |\varphi_j(x)|^2 + 2
\sumtwo{x\in\Lambda, y\notin\Lambda}{|x-y|=1} |\varphi_j(x)| \, |\varphi_j(y)|
& \text{if } 1\leq j\leq |\Lambda| \\ (4d+U) \sum_{x \in \Lambda} |\varphi_j(x)|^2 + 2
\sumtwo{x\in\Lambda, y\notin\Lambda}{|x-y|=1} |\varphi_j(x)| \, |\varphi_j(y)|
& \text{if } |\Lambda|<j\leq|\Omega|. \end{cases}
\end{equation}
Then \eqref{B5} takes the simpler form
\be
\sum_{j=1}^N e_j^U \geq \sum_{j=1}^{|\Lambda|} (e_j - \delta_j) + \sum_{j=|\Lambda|+1}^N (\bar e_j + U -
\delta_j).
\end{equation}
The sequence in the RHS is not necessarily increasing, but one gets a lower bound by
rearranging the terms. Hence one can apply Hardy,
Littlewood, P\'olya inequality. Indeed, it also works when the total sum over elements of the
sequences are not equal, provided the concave function is increasing --- which is the case with
$g(e)$. One obtains
\be
\sum_{j=1}^{|\Omega|} g(e_j^U) \geq \sum_{j=1}^{|\Lambda|} g(e_j - \delta_j) +
\sum_{j=|\Lambda|+1}^{|\Omega|} g(\bar e_j + U - \delta_j).
\end{equation}
We use now $g(e-\delta) \geq g(e) - \delta$, and we find
\be
F_{\Omega,\Lambda}^U(\beta,\mu) \geq F_\Lambda(\beta,\mu) +
F_{\Omega\setminus\Lambda}(\beta, \mu-U) - \sum_{j=1}^{|\Omega|} \delta_j.
\end{equation}

The remaining effort consists in estimating the sum of $\delta_j$, using
exponential decay of eigenfunctions $\varphi_j$ either in $\Lambda$ or in
$\Omega\setminus\Lambda$. Retracing \eqref{someupperbound} and \eqref{B6}, we get
\ba
\sum_{j=1}^{|\Lambda|} \delta_j &\leq 4d \, 2^d \eta(U) |\partial\Lambda| +
\frac{8d^2}{U-2d} |\partial\Lambda|, \nn\\
\sum_{j=|\Lambda|+1}^{|\Omega|} \delta_j &\leq (U+4d) \, 2^d \eta(U)
|\partial(\Omega\setminus\Lambda)| + \frac{8d^2}{U-2d} |\partial\Lambda|.
\end{align}
Notice that the last term can be written with $|\partial\Lambda|$ instead of
$|\partial(\Omega\setminus\Lambda)|$, as can be seen from \eqref{B6}. We use
$\partial(\Omega\setminus\Lambda)| \leq 2d |\partial\Lambda|$, and we finally obtain
\be
\sum_{j=1}^{|\Omega|} \delta_j \leq (2dU+4d+8d^2) \, 2^d \eta(U) |\partial\Lambda|
+ \frac{(4d)^2}{U-2d} |\partial\Lambda|.
\end{equation}
\end{proof}

\subsection{Upper bound}

We turn to the upper bound for the electronic free energy. We first notice that the free
energy is raised when one decorrelates the domain occupied by the classical particles,
from the empty domain. The following proposition applies to all finite subsets of $\bbZ^d$,
and it also applies when $\Omega$ is a finite $d$-dimensional torus.

\begin{proposition}
\label{propubtemp}
We have the upper bounds
\begin{itemize}
\item $F_{\Omega,\Lambda}^U(\beta,\mu) \leq F_\Lambda(\beta,\mu) +
F_{\Omega\setminus\Lambda}(\beta,\mu-U)$.
\item $F_\Lambda(\beta,\mu) \leq |\Lambda| f(\beta,\mu) + \frac1{1+\e{-\beta\mu}} \bigl( \frac{4\pi\sqrt d}{|S_d|^{1/d}}
|\Lambda|^{\frac{d-1}d} + 2d |\partial\Lambda| \bigl)$.
\end{itemize}
\end{proposition}

Notice that the isoperimetric inequality implies that for all finite $\Lambda\subset\bbZ^d$,
$|\Lambda|^{\frac{d-1}d} \leq |\partial\Lambda|$. This does not hold, however, when
$\Lambda$ is e.g.\ a box with periodic boundary conditions.

\begin{proof}
The Peierls inequality allows us to write
\be
\Tr \e{-\beta (H_{\Omega,\Lambda}^U - \mu N_\Omega)} \geq \sum_j \e{-\beta (\psi_j,
[H_{\Omega,\Lambda}^U - \mu N_\Omega] \psi_j)},
\end{equation}
for any set of orthonormal functions $\{\psi_j\}$ (in the Fock space of antisymmetric wave
functions on $\Omega$). We can choose the $\psi_j$ to be eigenfunctions of $H_\Lambda$
and $H_{\Omega\setminus\Lambda}$ --- decorrelating $\Lambda$ and
$\Omega\setminus\Lambda$. In $\Omega\setminus\Lambda$, the free electrons experience a
uniform potential $U$; the energy levels are given by the spectrum of
$h_{\Omega\setminus\Lambda}$ plus $U$. This only shifts the chemical potential, so that we
obtain the first claim of the proposition.

Now we estimate $F_\Lambda(\beta,\mu)$. Let us introduce
\be
\tilde e_j = (1-\tfrac{|\partial\Lambda|}{|\Lambda|}) e_j^* + 2d \tfrac{|\partial\Lambda|}{|\Lambda|} = \int\dd
\mu_j(k) \bigl[ \varepsilon_k + (2d-\varepsilon_k) \tfrac{|\partial\Lambda|}{|\Lambda|} \bigr];
\end{equation}
then $\tilde e_j \leq \tilde e_{j+1}$, $\sum_{j=1}^{|\Lambda|} \tilde e_j = \sum_{j=1}^{|\Lambda|} e_j^*$, and
the upper bound for the ground state energy can be cast in the form
\be
\sum_{j=1}^N e_j \leq \sum_{j=1}^N \tilde e_j.
\end{equation}

This allows us to summon again the Hardy, Littlewood, P\'olya inequality, and we get
\be
\label{B7}
F_\Lambda(\beta,\mu) \leq \sum_{j=1}^{|\Lambda|} g\Bigl( \int\dd\mu_j(k) \bigl[ \varepsilon_k + (2d-\varepsilon_k)
\tfrac{|\partial\Lambda|}{|\Lambda|} \bigr] \Bigr).
\end{equation}
The derivative of $g(e)$ satisfies
\be
0 < g'(e) \leq \frac1{1+\e{-\beta\mu}}
\end{equation}
(recall that $e\geq0$). Since the measure $\mu_j$ is concentrated on those $k$ where
$\varepsilon_k$ lies between $\varepsilon_{\rm F}(\tfrac{j-1}{|\Lambda|})$ and $\varepsilon_{\rm F}(\tfrac j{|\Lambda|})$, we
can bound \eqref{B7} by
\be
F_\Lambda(\beta,\mu) \leq \frac{|\Lambda|}{(2\pi)^d} \int_{[-\pi,\pi]^d} \dd k \,
g(\varepsilon_k) + \frac1{1+\e{-\beta\mu}} \Bigl( \sum_{j=1}^{|\Lambda|} \bigl[ \varepsilon_{\rm F}(\tfrac j{|\Lambda|}) - \varepsilon_{\rm
F}(\tfrac{j-1}{|\Lambda|}) \bigr] + 2d |\partial\Lambda| \Bigr).
\end{equation}

We need a bound for $\varepsilon_{\rm F}(\frac j{|\Lambda|}) - \varepsilon_{\rm F}(\frac{j-1}{|\Lambda|})$;
since $\nabla\varepsilon_k = 2(\sin k_1, \dots, \sin k_d)$, we have
$\|\nabla\varepsilon_k\| \leq 2\sqrt d$. Let us take $k$ such that $\varepsilon_k =
\varepsilon_{\rm F}(\frac{j-1}{|\Lambda|})$, and $\delta k \parallel k$ such that
$\varepsilon_{k+\delta k} = \varepsilon_{\rm F}(\tfrac j{|\Lambda|})$. Then
\be
\label{boundFermi}
\varepsilon_{\rm F}(\tfrac j{|\Lambda|}) - \varepsilon_{\rm F}(\tfrac{j-1}{|\Lambda|}) \leq 2\sqrt d \, \|\delta k\|.
\end{equation}
If $\delta k_{\rm min}$ is chosen so as to minimize the norm of such $\delta k$, we have
\be
\frac1{|\Lambda|} = \frac1{(2\pi)^d} \int_{\varepsilon_{\rm F}(\frac{j-1}{|\Lambda|}) < \varepsilon_k < \varepsilon_{\rm
F}(\frac j{|\Lambda|})} \dd k \geq \frac1{(2\pi)^d} \|\delta_{\rm min}\|^d |S_d|.
\end{equation}
Combining this inequality with \eqref{boundFermi}, we get
\be
\varepsilon_{\rm F}(\tfrac j{|\Lambda|}) - \varepsilon_{\rm F}(\tfrac{j-1}{|\Lambda|}) \leq \frac{4\pi\sqrt
d}{|S_d|^{1/d}} |\Lambda|^{-1/d}.
\end{equation}
This leads to the upper bound of Proposition \ref{propubtemp}.
\end{proof}

\subsection{Proofs of the corollaries}

\begin{proof}[Proof of Corollary \ref{corIsing}]
Let $e^- = -d+h$ and $e^+ = -d-h$ be the energies per site of the all $-$ and all $+$ Ising
configurations. A configuration can be specified by the set $\Lambda$ of $-$ spins. Let
$\caB(\Lambda)$ be the set of bonds connecting $\Lambda$ and
$\Omega\setminus\Lambda$. Notice that $\frac1{2d} |\caB(\Lambda)| \leq |\partial\Lambda| \leq
|\caB(\Lambda)|$. The partition function of the Ising model can be written as
\be
Z_{\text I, \Omega} = \sum_{\Lambda\subset\Omega} \e{-\beta [|\Lambda| e^- +
|\Omega\setminus\Lambda| e^+]} \e{-2\beta |\caB(\Lambda)|}.
\end{equation}
Now the upper bound for $F_{\Omega,\Lambda}^U(\beta,\mu)$ implies that the partition
function of the Falicov-Kimball model is bounded below by
\be
Z_\Omega \geq \sum_{\Lambda\subset\Omega} \e{-\beta [|\Lambda| f(\beta,\mu) +
|\Omega\setminus\Lambda| f(\beta,\mu-U)]} \e{-\beta C_{d,\mu} |\partial\Lambda|} \e{-\beta
C_{d,\mu}' |\Omega|^{\frac{d-1}d}}.
\end{equation}
The last factor vanishes in the thermodynamic limit. One then makes the connection with
Ising by multiplying $Z_\Omega$ by
$$
\exp\Bigl\{ \beta |\Omega| \Bigl(\tfrac d2 C_{d,\mu} + \tfrac12 \bigl[ f(\beta,\mu) +
f(\beta,\mu-U) \bigr] \Bigr) \Bigr\},
$$
and by choosing the temperature to be $\frac12 C_{d,\mu} \beta$, and the magnetic field
to be
\be
h = \frac1{C_{d,\mu}} [f(\beta,\mu) - f(\beta,\mu-U)]
\end{equation}
(the magnetic field is negative). The other bound is similar, simply replace $C_{d,\mu}$
by $\bar\alpha/2d$.
\end{proof}

\begin{proof}[Proof of Corollary \ref{coreqst}]
Because $\Omega$ was assigned periodic boundary conditions, we have
\be
\expval{\delta_{w_x,w_y}}_\Omega^{\phantom{x}} = \frac1{|\Omega|} \bigexpval{\sum_{z\in\Omega}
\delta_{w_{x+z},w_{y+z}}}_\Omega.
\end{equation}
It is not hard to check that for any configuration $w$ specified by $\Lambda \subset
\Omega$, one has
\be
\sum_{z\in\Omega} \delta_{w_{x+z},w_{y+z}} \geq |\Omega| - |\partial\Lambda| \, |x-y|_1.
\end{equation}
Then
\be
\expval{\delta_{w_x,w_y}}_\Omega^{\phantom{x}} \geq 1 - |x-y|_1
\bigexpval{\frac{|\partial\Lambda|}{|\Omega|}}_\Omega.
\end{equation}
We need a bound for the last term. The fact is that typical configurations of classical
particles cannot have too much boundary: $\frac{|\partial\Lambda|}{|\Omega|}$ is smaller
than $r = \frac{2\log2}{\beta\bar\alpha(\beta,\mu)}$. Indeed,
\ba
&\frac{\sum_{\Lambda\subset\Omega} \charfct{|\partial\Lambda| > r |\Omega|} \e{-\beta
F_{\Omega,\Lambda}^U(\beta,\mu)}}{\sum_{\Lambda\subset\Omega}
\e{-\beta F_{\Omega,\Lambda}^U(\beta,\mu)}} \nn\\
&\hspace{2cm} \leq \frac{2^{|\Omega|} \e{-\beta |\Lambda|
f(\beta,\mu) - \beta (|\Omega|-|\Lambda|) f(\beta,\mu-U)} \e{-(2\log2) |\Omega|}}{\e{-\beta |\Lambda|
f(\beta,\mu) - \beta (|\Omega|-|\Lambda|) f(\beta,\mu-U)} \e{-\beta C_{d,\mu} (n_{\rm c}
|\Omega|)^{\frac{d-1}d}} \e{-\beta C_{d,\mu}' |\Omega|^{\frac{d-1}d}}} \\
&\hspace{2cm} \leq 2^{-|\Omega|} \e{\beta (C_{d,\mu} + C_{d,\mu}') |\Omega|^{\frac{d-1}d}}.
\nn
\end{align}
Therefore
\be
\expval{\delta_{w_x,w_y}}_\Omega^{\phantom{x}} \geq 1 - \frac{2\log2}{\beta\bar\alpha(\beta,\mu)}
|x-y|_1 - 2^{-|\Omega|} \e{\beta (C_{d,\mu} + C_{d,\mu}') |\Omega|^{\frac{d-1}d}}.
\end{equation}
The last term vanishes in the limit $\Omega \nearrow \bbZ^d$, and the term involving $|x-y|_1$
vanishes when $\beta\to\infty$.
\end{proof}

\section{Conclusion}
\label{secconcl}

Our analysis of the Falicov-Kimball model away from half-filling allows some
extrapolations. We expect segregation to survive at small temperature, when both the
classical particles and the electrons are described by the grand-canonical ensemble, at
inverse temperature $\beta$ and with chemical potentials $\mu_{\rm c}$ and $\mu_{\rm e}$.
Segregation is a manifestation of coexistence between a phase with many classical particles
and few electrons, and a phase with many electrons and few classical particles. It is
therefore natural to conjecture the following, for $d\geq2$:

\begin{quote}{\it
A first order phase transition occurs at low temperature, when varying the chemical
potentials.}
\end{quote}

The transition from the chessboard state at half-filling for the itinerant and
heavy electrons (and large $U$) to the segregated
state is still not clear. A heuristic analysis suggests that these states could coexist,
hence there could be another first-order phase transition. Alternate possibilities
include mixtures between other periodic phases and the empty or full lattice
before the segregation sets in.

One interest of the Falicov-Kimball model is its possible relevance in understanding the
Hubbard model, a notoriously difficult task. See e.g.\ \cite{Lieb2} and \cite{Tas2} for
reviews of
rigorous results on the Hubbard model. The relationship between
the Falicov-Kimball model and the Hubbard model is like the one between the Ising and
Heisenberg models for magnetism. The former does not possess the continuous symmetry of the latter, and
therefore the approximation is a crude one. Still, the two models share many similarities;
for instance, the Falicov-Kimball model displays long-range order of the chessboard type at
half-filling and at low temperature \cite{KL}, and the ground state of the Hubbard model is
a spin singlet \cite{Lieb}.

Ferromagnetism in the Hubbard model
depends on the dimension, on the filling, and on the geometry: it has
been shown to occur on special lattices such as `line-graphs' \cite{MT, Tas, Mie, Tas2}.
Does ferromagnetism take place in the Hubbard model on $\bbZ^3$, for large
repulsions and away from half-filling?

Returning to Falicov-Kimball, let us walk on the road that leads to Hubbard. We consider the
{\it asymmetric Hubbard model} that describes spin $\frac12$ electrons with hoppings
depending on the spins (this interpretation is more convenient than physical). Its
Hamiltonian is
\be
H_t = -\sum_{x,y: |x-y|=1} c_{x\uparrow}^\dagger c_{y\uparrow} - t \sum_{x,y:
|x-y|=1} c_{x\downarrow}^\dagger c_{y\downarrow} + U \sum_x n_{x\uparrow} n_{x\downarrow}.
\end{equation}
Notice that $H_0$ is the Falicov-Kimball model, while $H_1$ is the usual Hubbard model.
Although we did not prove it, it is rather clear that segregation still takes place for
very small $t$. Furthermore, the density of the phase with classical particles, in the
ground state, should still be exactly 1 --- indeed, the electrons exert a sort of
`pressure' that packs the classical particles together, and the tendency of the latter to delocalize is
not strong enough to overcome this pressure. This is summarized in the following
conjecture:

\begin{quote}{\it
For $t\leq t_0$, segregation occurs in the ground state, at large $U$ and away from
half-filling, in the form of a coexistence between a phase of classical particles with
density 1, and a phase of electrons with smaller density.}
\end{quote}

This should also hold at positive temperature, although the density of the phase of the classical
particles will be reduced, due to the presence of some holes.

If we increase $t$, assuming that segregation remains, we should reach a critical value
$t_{\rm c}<1$ where the region of classical particles starts to grow. The density of the
phase of particles with smaller hoppings is now strictly less than 1. A major question is
whether segregation survives all the way to the point where $t$ reaches 1 --- this would imply the
existence of a ferromagnetic phase in the Hubbard model. We note, however, that while it is
conceivable that there is a segregated (i.e., ferromagnetic) ground state at $t=1$, it
cannot be true that {\it every} ground state (for equal number of up and down spins) is
segregated. This follows from the SU(2) symmetry. If $\Psi$ is a saturated ferromagnetic
ground state with $2N$ up electrons, we can construct $\Phi = (S_-)^N \Psi$, which is
also a ground state, with $N$ up and $N$ down electrons. However, $\Phi$ has the up and down
electrons inextricably mixed, which is the opposite of a segregated state.
Indeed, the SU(2) symmetry is restored precisely at t=1, and the ground states
have at least the degeneracy due to this symmetry.

The Hubbard model is a rich and complicated model that poses difficult challenges. The
Falicov-Kimball model can be of some help, for instance in checking scenarios that should
apply to both models. This discussion of ferromagnetism illustrates however that
the links between them are subtle.

\appendix
\renewcommand{\theequation}{\arabic{equation}}
\numberwithin{equation}{section}

\section{}

We derive in the sequel various expressions that are too intricate to appear in the main
body of this paper.

\begin{lemma}
\label{lemapp}
\hfill
\begin{itemize}
\item[(a)] $\frac1{(2\pi)^d} \int\dd k |\varepsilon_k - \varepsilon_{\rm F}|^{-1/4} < 2$.
\item[(b)] Assume that $\alpha^2 \leq \frac{16 \sqrt2 \pi d}{|S_d|^{1/d}} n^{1/d}$; then for
all $k$ such that $\varepsilon_k = \varepsilon_{\rm F}$,
$$
\int_{\varepsilon_{k'} < \varepsilon_{\rm F}} \dd k' \, \charfct{|k'-k|<\alpha} \geq |S_d|
\bigl( \frac{\alpha^2}{8\pi d} \bigr)^d.
$$
\item[(c)] $e(n) \geq 12 (\frac9{10})^d n^{1+\frac2d} / |S_d|^{\frac2d}$.
\item[(d)] $\| \nabla \frac{\|b_k\|^2}{|\partial\Lambda|} \| \leq 8d^{5/2}$.
\item[(e)] $\| \nabla \frac{\| (h_\Lambda-\varepsilon) b_k \|^2}{|\partial\Lambda|} \| \leq
2^9 d^{11/2}$.
\item[(f)] $\| \nabla \frac{(b_k, h_\Lambda b_k)}{|\partial\Lambda|} \| \leq
32 d^{7/2}$.
\item[(g)] Assume that $\|b_k\|^2/|\partial\Lambda| \geq \eta$. Then if $\eta\leq1$, $\|
\nabla \frac{(b_k, h_\Lambda b_k)}{\|b_k\|^2} \| \leq \eta^{-2} 2^8 d^{11/2}$.
\end{itemize}
\end{lemma}

\begin{proof}[Proof of Lemma \ref{lemapp} (a)]
Setting $Y = 2d - \varepsilon_{\rm F} - 2\sum_{i=2}^d \cos k_i$, and making the change of
variables $\xi = \cos k_1$, one gets
\ba
\int\dd k |\varepsilon_k - &\varepsilon_{\rm F}|^{-1/4} = 2 \int_{[-\pi,\pi]^{d-1}} \dd k_2
\dots \dd k_d \int_0^\pi \dd k_1 \frac1{|Y - 2\cos k_1|^{1/4}} \nn\\
&= 2 \int_{[-\pi,\pi]^{d-1}} \dd k_2
\dots \dd k_d \int_{-1}^1 \dd\xi \frac1{\sqrt{1-\xi^2}} \frac1{|Y - 2\xi|^{1/4}} \nn\\
&\leq 2 \int_{[-\pi,\pi]^{d-1}} \dd k_2 \dots \dd k_d \Bigl( 2 \int_0^1 \dd\xi
\frac1{(1-\xi^2)^{3/4}} \Bigr)^{2/3} \Bigl( \int_{-1}^1 \dd\xi |Y - 2\xi|^{-3/4}
\Bigr)^{1/3} \nn\\
&\leq 2 \int_{[-\pi,\pi]^{d-1}} \dd k_2 \dots \dd k_d \Bigl( 2 \int_0^1 \dd\zeta
\frac1{\sqrt\zeta (1-\zeta)^{3/4}} \Bigr)^{2/3} \Bigl( 2^{-3/4} \int_{-1}^1 \dd\xi |\xi|^{-3/4}
\Bigr)^{1/3} \nn
\end{align}
The integral over $\zeta$ can be split into one running from 0 to $\frac12$, and one
running from $\frac12$ to 1. For the first part we bound $\frac1{\sqrt\zeta
(1-\zeta)^{3/4}} \leq 2^{3/4} \frac1{\sqrt\zeta}$, while the bound for the second part can
be chosen to be $\sqrt2 \frac1{(1-\zeta)^{3/4}}$. Everything can now be computed
explicitly, and we find $2^{31/12} 3^{2/3} (2\pi)^{d-1} < 2(2\pi)^d$.
\end{proof}

\begin{proof}[Proof of Lemma \ref{lemapp} (b)]
Let us introduce a map $\gamma(\xi)$ such that $1-\frac12 \gamma^2(\xi) = \cos\xi$;
precisely,
\be
\gamma(\xi) = \begin{cases} \phantom- \sqrt{2(1-\cos\xi)} & \text{if } \xi \in [0,\pi] \\
-\sqrt{2(1-\cos\xi)} & \text{if } \xi \in [-\pi,0]. \end{cases}
\end{equation}
The condition $\varepsilon_k < \varepsilon_{\rm F}$ becomes $\sum_{i=1}^d |\gamma(k_i)|^2 <
\varepsilon_{\rm F}$. The derivative of $\gamma$ is
\be
\frac{\dd\gamma}{\dd\xi} = \frac{|\sin\xi|}{\sqrt{2(1-\cos\xi)}}.
\end{equation}

We check now that $|\gamma(\xi')-\gamma(\xi)| > |\xi'-\xi|^2/4\pi$. Let us assume that
$\gamma(\xi')>\gamma(\xi)$. Then
\ba
\gamma(\xi')-\gamma(\xi) &= \int_\xi^{\xi'} \dd\lambda
\frac{|\sin\lambda|}{\sqrt{2(1-\cos\lambda)}} \geq \frac12 \int_\xi^{\xi'} \dd\lambda
|\sin\lambda| \\
&\geq \frac1\pi \int_\xi^{\xi'} \dd\lambda |\lambda| \geq |\xi'-\xi|^2/4\pi.
\end{align}

Then we can write
\ba
\int_{\varepsilon_{k'} < \varepsilon_{\rm F}} \dd k' \, \charfct{|k'-k|<\alpha} &\geq
\int_{\varepsilon_{k'} < \varepsilon_{\rm F}} \dd k' \, \charfct{|k_i'-k_i| <
\tfrac\alpha{\sqrt d} \; \forall i} \nn\\
&\geq \int\dd\gamma_1' \dots \dd\gamma_d' \,\charfct{\sum_{i=1}^d |\gamma_i'|^2 <
\varepsilon_{\rm F}} \,\charfct{|\gamma_i'-\gamma_i| < \tfrac{\alpha^2}{4\pi d}} \nn
\end{align}
One gets a lower bound by replacing the last characteristic function by the condition
$\sum_{i=1}^d |\gamma_i'-\gamma_i|^2 < \frac{\alpha^2}{4\pi d}$. Recall that
$\varepsilon_{\rm F} \geq \frac{32}{|S_d|^{2/d}} n^{2/d}$; the assumption of the
lemma implies that $\sqrt{\varepsilon_{\rm F}} >
\frac{\alpha^2}{4\pi d}$; as a consequence, a lower bound is the volume of the sphere of
radius $\frac{\alpha^2}{8\pi d}$.
\end{proof}

\begin{proof}[Proof of Lemma \ref{lemapp} (c)]
By \eqref{boundsepsk},
\be
e(n) \geq \frac8{\pi^2 (2\pi)^d} \int_{|k|^2 < \varepsilon_{\rm F}} \dd k |k|^2 = \frac{8d
|S_d|}{\pi^2 (2\pi)^d (d+2)} \varepsilon_{\rm F}^{\frac d2+1}.
\end{equation}
The lower bound then follows from
\be
\varepsilon_{\rm F} \geq \frac{32}{|S_d|^{2/d}} n^{2/d}.
\end{equation}
\end{proof}

\begin{proof}[Proof of Lemma \ref{lemapp} (d)--(g)]
Since
\be
\|b_k\| = \sum_{x\in\partial\Lambda} \sumtwo{e: x+e \notin \Lambda}{e': x+e' \notin \Lambda}
\e{\ii k(e-e')},
\end{equation}
we have
\be
\bigl( \nabla \frac{\|b_k\|}{|\partial\Lambda|} \bigr)_j = \frac1{|\partial\Lambda|}
\sum_{x\in\partial\Lambda} \sum_{e,e'} \ii (e_j-e_j') \e{\ii k(e-e')}.
\end{equation}
This is less than $2(2d)^2$, and we obtain the bound (d).

We consider now $\|(h_\Lambda-\varepsilon) b_k\|^2$.
\ba
\e{\ii kx} [(h_\Lambda-\varepsilon) b_k](x) &= (2d-\varepsilon) \sum_{e': x+e' \notin
\Lambda} \e{-\ii ke'} - \sumtwo{e: x+e \in \partial\Lambda}{e': x+e+e' \notin \Lambda}
\e{-\ii k(e+e')} \nn\\
&= \sumtwo{e: x+e \in \partial\Lambda, |e|=0,1}{e': x+e+e' \notin \Lambda} \e{-\ii k(e+e')}
\bigl( (2d-\varepsilon) \charfct{|e|=0} - \charfct{|e|=1} \bigr).
\end{align}
In the last line, $e$ is allowed to be 0. Let $\xi(e) = \bigl( (2d-\varepsilon)
\charfct{|e|=0} - \charfct{|e|=1} \bigr)$. Then
\be
\bigl| \bigl[ (h_\Lambda-\varepsilon) b_k \bigr](x) \bigr|^2 = \sumtwo{e: x+e \in
\partial\Lambda, |e|=0,1}{e': x+e+e' \notin \Lambda} \; \sumtwo{e'': x+e'' \in
\partial\Lambda, |e''|=0,1}{e''': x+e''+e''' \notin \Lambda} \e{\ii k(e+e'-e''-e''')} \xi(e)
\xi(e'').
\end{equation}
One computes now the $j$-th component of the gradient; it involves a term $e_j + e_j' -
e_j'' - e_j'''$ that is smaller than 4; there are sums over $e',e'''$, with less than $(2d)^2$ terms;
the sum $\sum_e |\xi(e)|$ is bounded by $4d$; finally, the number of sites where
$(h_\Lambda-\varepsilon) b_k$ differs from 0 is bounded by $2d|\partial\Lambda|$. As a
result, the $j$-th component of the gradient is bounded by $\frac12 (4d)^5$, and we obtain
(e).

We estimate now the gradient of $(b_k, h_\Lambda b_k)$. One easily checks that
\be
(b_k, h_\Lambda b_k) = \|b_k\|^2 - \sum_{x\in\partial\Lambda} \sum_{e: x+e
\notin\Lambda} \sum_{e': x+e' \in \partial\Lambda} \sum_{e'': x+e'+e'' \notin \Lambda} \e{\ii
k(e-e'-e'')}.
\end{equation}
We can use the bound (d) for the gradient of $\|b_k\|^2$. The gradient of the last term is less than $3(2d)^3
|\partial\Lambda|$, so we can write
\be
\Bigl\| \nabla \frac{(b_k, h_\Lambda b_k)}{|\partial\Lambda|} \Bigr\| \leq 8d^{5/2} + 24
d^{7/2} \leq 32 d^{7/2}.
\end{equation}

Finally, one easily checks that
\be
\bigl\| \nabla \frac{(b_k, h_\Lambda b_k)}{\|b_k\|^2} \Bigr\|^2 \leq 2 \Bigl(
\frac{|\partial\Lambda|}{\|b_k\|^2} \Bigr)^2 \Bigl\| \nabla \frac{(b_k, h_\Lambda
b_k)}{|\partial\Lambda|} \Bigr\|^2 + 2 \Bigl( \frac{(b_k, h_\Lambda
b_k)}{|\partial\Lambda|} \Bigr)^2 \Bigl(
\frac{|\partial\Lambda|}{\|b_k\|^2} \Bigr)^4 \Bigl\| \nabla
\frac{\|b_k\|^2}{|\partial\Lambda|} \Bigr\|^2.
\end{equation}
Using (d) and (f), as well as $(b_k, h_\Lambda b_k)/|\partial\Lambda| \leq 2(2d)^3$, one
gets (g).
\end{proof}

\vspace{3mm}
\noindent
{\it Acknowledgments:}
It is a pleasure to thank Michael Loss for several valuable discussions.

\end{document}